\documentclass[showpacs,preprintnumbers,amsmath,amssymb,prb]{revtex4}

\def\XXint#1#2#3{{\setbox0=\hbox{$#1{#2#3}{\int}$}
     \vcenter{\hbox{$#2#3$}}\kern-.5\wd0}}

\usepackage{graphicx}
\usepackage{subfigure}
\usepackage{bm}

\begin{document}

\title{Spiral magnets with Dzyaloshinskii-Moriya interaction containing defect bonds}

\author{O.\ I.\ Utesov$^1$}
\email{utiosov@gmail.com}
\author{A.\ V.\ Sizanov$^{1,2}$}
\email{alexey.sizanov@gmail.com}
\author{A.\ V.\ Syromyatnikov$^{1,2}$}
\email{asyromyatnikov@yandex.ru}
\affiliation{$^1$National Research Center "Kurchatov Institute" B.P.\ Konstantinov Petersburg Nuclear Physics Institute, Gatchina 188300, Russia}
\affiliation{$^2$Department of Physics, Saint Petersburg State University, Ulianovskaya 1, St.\ Petersburg 198504, Russia}

\date{\today}

\begin{abstract}

We present a theory describing spiral magnets with Dzyaloshinskii-Moriya interaction (DMI) subject to bond disorder at small concentration $c$ of defects. It is assumed that both DMI and exchange coupling are changed on imperfect bonds. Qualitatively the same physical picture is obtained in two models which are considered in detail: B20 cubic helimagnets and layered magnets in which DMI leads to a long-period spiral ordering perpendicular to layers. We find that the distortion of the spiral magnetic ordering around a single imperfect bond is long-range: values of additional turns of spins decay with the distance $r$ to the defect as $1/r^2$ being governed by the Poisson's equation for electric dipole. At finite concentration of randomly distributed imperfect bonds, we calculate correction to the spiral vector. We show that this correction can change the sign of spin chirality even at $c\ll1$ if defects are strong enough. It is demonstrated that impurities lead to a diffuse elastic neutron scattering which has power-law singularities at magnetic Bragg peaks positions. Then, each Bragg peak acquires power-law decaying tails. Corrections are calculated to the magnon energy and to its damping caused by scattering on impurities.

\end{abstract}


\maketitle

\section{Introduction}

In crystals without center of inversion, Dzyaloshinskii-Moriya interaction (DMI) is caused by an antisymmetric spin-orbit interaction. \cite{dzyal1958, moriya1960} The competition of the symmetric ferromagnetic (FM) or antiferromagnetic (AF) exchange interaction and DMI can result in a spiral magnetic structure. \cite{dzyal1964} Although a long time has passed since the spiral ordering was observed for the first time, helimagnets with DMI still attract a lot of attention. This interest is stimulated by discovery of rich phase diagrams and exotic spin structures caused by DMI which arise under certain conditions. Phases with such topological states as chiral soliton lattices in layered helimagnets (e.g., in $\rm Cr_{1/3} Nb S_2$) \cite{togawa} and skyrmion lattices in B20 cubic chiral magnets (e.g., in MnSi) \cite{muhlbauer} are widely discussed now. These materials are attractive not only from a fundamental but also from a technological point of view owing to their potential applications in spintronic devices.

Mixed B20 spiral compounds have been considered experimentally recently. \cite{grig2013} It is shown in Ref.~\cite{grig2013} that the modulus of the spiral vector $\bf q$ in Mn$_{1-x}$Fe$_x$Ge depends on dopant concentration $x$ and the magnetic chirality changes its sign (and $\bf q$ goes through zero) at $x\approx0.75$. This observation is quite expected because MnGe and FeGe are B20 cubic helimagnets with opposite signs of the spin chirality. Evidently, such a behavior is a consequence of the fact that the exchange interaction and DMI change around dopant ions which can be considered as defects at $x\ll1$ or $x\approx1$. These experimental results are interpreted phenomenologically by renormalization of constants in the Hamiltonian describing the pure translationally invariant B20 magnets. Then, a more detailed theoretical description of mixed spiral materials is required.

\begin{figure}
  \noindent
  \includegraphics[scale=0.65]{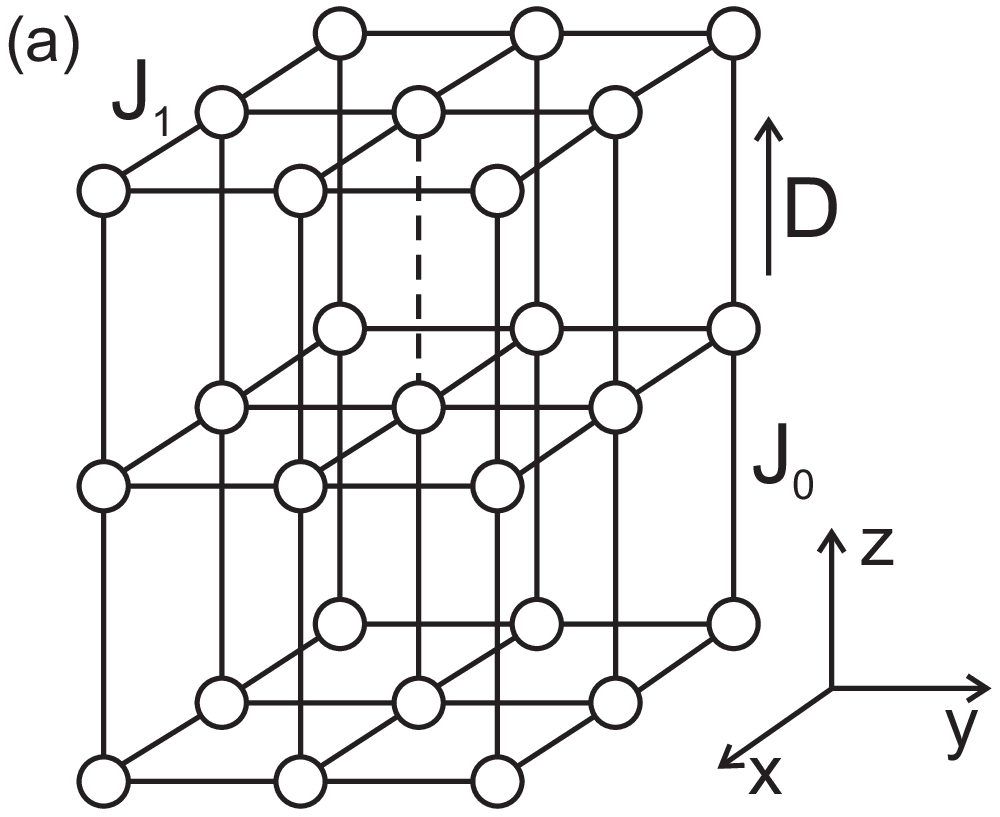}
  \includegraphics[scale=0.65]{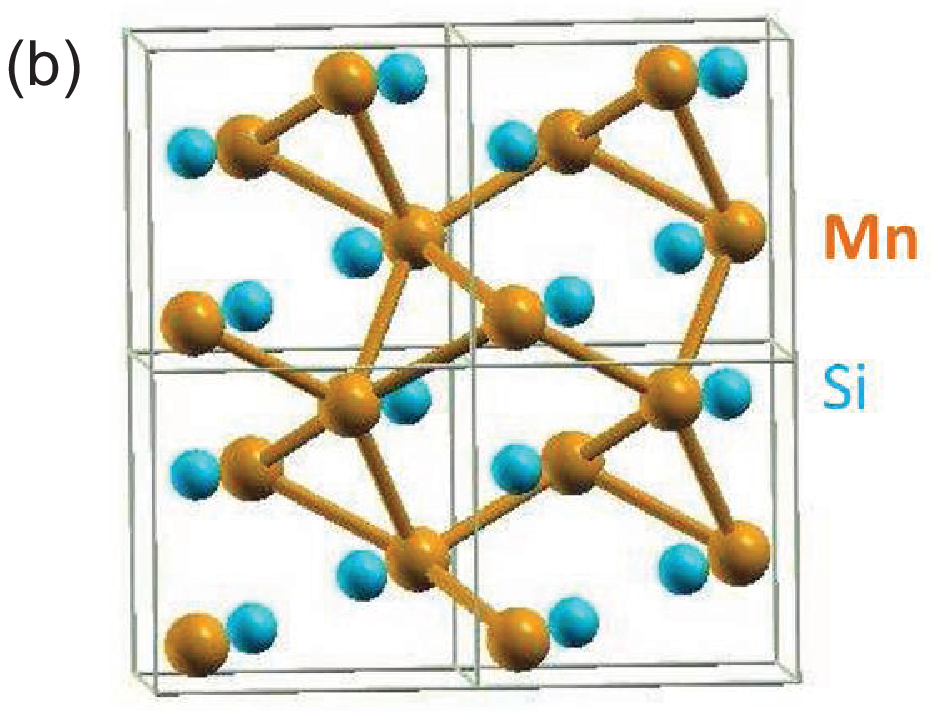}
  \hfil
  \caption{(Color online.)
Two types of spiral magnets with DMI considered in the present paper.
(a) Layered spiral magnet \eqref{H0start} with tetragonal lattice in which DMI acts only between nearest-neighbor spins from neighboring $xy$ planes (DMI vector $\bf D$ is depicted). Exchange coupling constants between neighboring spins inside $xy$ planes ($J_1$) and along $z$ axis ($J_0$) are also shown. The long-period helix propagates along $z$ axis. Similar models with hexagonal $xy$ planes describe $\rm Cr_{1/3}NbS_2$ and $\rm CsCuCl_3$. The imperfect bond is shown by dashed line.
	(b) Cristal structure of MnSi that is probably the most famous representative of spiral cubic B20 magnets. The helix can propagate along any space diagonal of the cube.}
  \label{Zmagnet}
\end{figure}

Motivated by this experimental activity, we address in the present paper the problem of spiral magnets with DMI subject to bond disorder at small concentration $c$ of defects. We assume that both exchange interaction and DMI are changed on imperfect bonds. Two models are considered in detail: i) spiral magnets on a simple cubic lattice with FM exchange coupling and small DMI between nearest-neighbor spins, where the DMI vector is directed along the line connected couple of spins, and ii) layered magnets with small DMI which acts between nearest-neighbor spins from different layers and which vector is directed along $z$ (chiral) axis perpendicular to layers (see Fig.~\ref{Zmagnet}). The most famous and the most studied compounds described by the model of the first type is probably MnSi and those of the second type are $\rm Cr_{1/3} Nb S_2$ and $\rm CsCuCl_3$. At zero magnetic field and small temperature, DMI leads to long-period helix structures in these materials along one of the space cubic diagonals and along $z$ axis, correspondingly. \cite{ishikawa1976,moriya1982,miyadai1983}

We obtain qualitatively the same physical picture in both models. The one-impurity problem is addressed first. We show that the perturbation of the spiral ordering around the defect bond (i.e., values of additional turns of spins due to the defect) is described by the Poisson's equation for electric dipole.
\footnote{It should be noted that this fruitful electrostatic analogy is not new in physics of mixed magnets. It was discovered first by Villain \cite{vill1,vill2} in a spin-glass problem. Later, the electrostatic analogy was used also in discussion of $\rm La_{2-\mathit x}Sr_{\mathit x}CuO_4$. \cite{aharony,korenblit}}
Then, the magnetic ordering disturbance made by one impurity is long-range: values of additional turns of spins decay with the distance $r$ to the defect as $1/r^2$. This finding can be easily extended to the corresponding models on lattices with space dimensions $d\ge2$, the result being $1/r^{d-1}$. The spin texture around a ferromagnetic bond observed in two-dimensional collinear AFs follows the same law (see Refs.~\cite{aharony,korenblit} and references therein). It has been found recently that the distortion of magnetic ordering around defects decays exponentially in collinear AFs in magnetic field \cite{henley,eggert,wollny2} while the spin texture around the vacancy in triangular AF decay as $1/r^{d+1}$. \cite{wollny1}

At finite defects concentration $c \ll 1$, spiral magnets we discuss are equivalent to a dielectric with randomly distributed electric dipoles which lead to a finite average ``polarization'' of a unit volume proportional to $c$. This ``electrical polarization'' corresponds to a correction $\delta q\propto c$ to the modulus of the spiral vector ${\bf q}$.

Our analysis of the elastic neutron scattering cross section predicts magnetic Bragg peaks (satellites) on momenta transfer ${\mbox{\boldmath $\cal Q$}}=\pm (\mathbf{q}+\delta{\bf q})+\mbox{\boldmath $\tau$}$, where $\mbox{\boldmath $\tau$}$ is a reciprocal lattice vector. Besides, we obtain a diffuse scattering. Quite unexpectedly for diffuse scattering caused by disorder, its cross section has power-law singularities at positions of magnetic Bragg peaks. This feature is attributed to the long-range character of the perturbation made by defect bonds. Thus, impurities result in the shift by $\delta{\bf q}$ of the magnetic Bragg peaks positions and in power-law decaying tails of each peak.

We calculate also magnon spectrum renormalization due to the scattering on defects in the first order in $c$. These calculations are performed in the layered helimagnets only for FM exchange coupling constants.

The rest of the present paper is organized as follows. Secs.~\ref{layer} and \ref{B20magnets} which have similar structures are devoted to layered helimagnets and to B20 cubic spiral magnets with DMI, respectively. In subsections \ref{layer}A and \ref{B20magnets}A, we consider Hamiltonians of pure systems using the conventional Holstein-Primakoff spin transformation. We discuss in subsections \ref{layer}B and \ref{B20magnets}B the perturbation of the magnetic ordering around one imperfect bond and consider small concentration of such bonds. Then, we present our results for elastic neutron scattering cross-section in systems with bond disorder (Secs.~\ref{layer}C and \ref{B20magnets}C). In Secs.~\ref{layer}D and \ref{B20magnets}D, magnon spectrum renormalization is considered. All calculations in Secs.~\ref{layer}A--\ref{layer}D are carried out for layered helimagnets with FM exchange interaction. We show in Sec.~\ref{otherlay} that these results (except for the spectrum renormalization) are applicable after simple modifications to many other layered helimagnets with bond disorder. Sec.~\ref{sum} contains the summary and the conclusion. One appendix is added with some details of the magnon spectrum calculation.

\section{Layered spiral magnets with DMI}
\label{layer}

\subsection{Pure system}

In this section we consider a magnet containing FM $xy$ planes with a simple square lattice and the exchange coupling between neighboring spins only. Planes are stacked along $z$ axis. We take into account the exchange coupling and DMI between neighboring spins from neighboring planes. The DMI vector ${\bf D}=D{\bf e}_z$ is the same for all bonds along $z$ axis, where ${\bf e}_z=(0,0,1)$ is the unit vector directed along $z$ axis and we assume for simplicity that the distance between all neighboring sites is equal to unity (see Fig.~\ref{Zmagnet}(a)). The Hamiltonian of this system has the form
\begin{equation}
  \label{H0start}
  \mathcal{H}_0=-J_0 \sum_{in} {\bf S}_{in} {\bf S}_{in+1}-J_1 \sum_{\langle ij\rangle n} {\bf S}_{in} {\bf S}_{jn}   -\sum_{in} {\bf D}\cdot \left[ {\bf S}_{in}\times {\bf S}_{in+1} \right],
\end{equation}
where $J_0,J_1>0$, $J_0,J_1 \gg D$, ${\bf S}_{in}$ is an operator of the spin sitting at $i$-th site of $n$-th plane, and $\langle ij\rangle n$ denote nearest neighbor sites in the $n$-th plane. The last term in Eq.~\eqref{H0start} containing antisymmetric combinations of spins $[{\bf S}_{in}\times {\bf S}_{in+1}]$ can be eliminated by applying the rotation about $z$ axis by a pitch $q$: \cite{nikuni1993}
\begin{eqnarray}
  \label{transform}
	S^{x}_{in} &=& S^{x^\prime}_{in} \cos{n q} - S^{y^\prime}_{in} \sin{n q},\nonumber\\
  S^{y}_{in} &=& S^{x^\prime}_{in} \sin{n q} + S^{y^\prime}_{in} \cos{n q},\\
	S^{z}_{in} &=& S^{z^\prime}_{in}.\nonumber
\end{eqnarray}
The value of $q$ is chosen so that the antisymmetric spin combinations disappear in the Hamiltonian. Simple calculation give
\begin{equation}
  \label{pitch}
  \tan q = \frac{D}{J_0}\ll1.
\end{equation}
After transformation \eqref{transform}, Hamiltonian \eqref{H0start} obeys the following form:
\begin{equation}
  \label{H0final}
  \mathcal{H}=-\sum_{in} \left[ J_0 S^{z'}_{in} S^{z'}_{in+1} + \tilde{J}_0 (S^{x^\prime}_{in} S^{x^\prime}_{in+1}+S^{y^\prime}_{in} S^{y^\prime}_{in+1}) \right]    -J_1 \sum_{\langle ij\rangle n} {\bf S}^{\prime}_{in} {\bf S}^{\prime}_{jn},
\end{equation}
where $\tilde{J}_0= J_0\sqrt{1+(D/J_0)^2}$ and ${\bf S}^{\prime}_{in}=(S^{x'}_{in},S^{y'}_{in},S^{z'}_{in})$. Thus, the initial Hamiltonian \eqref{H0start} of the system with the spiral spin ordering described by the vector ${\bf q}=(0,0,q)$, where $q$ is given by Eq.~\eqref{pitch}, is equivalent to a FM described by Hamiltonian \eqref{H0final}. As $\tilde{J}_0>J_0$, $xy$ plane is the easy one in FM \eqref{H0final}. Then, DMI forces spins to lie in the plane perpendicular to $\bf D$.

For further consideration of Hamiltonian \eqref{H0final}, we use the Holstein-Primakoff spin representation
\begin{eqnarray}
  \label{SpinRep}
	S^{x^{\prime}}_{in}&=&S-a^{+}_{in} a_{in}, \nonumber \\
  S^{y^{\prime}}_{in}&\approx&\sqrt{\frac{S}{2}}\left(a^{+}_{in}+a_{in}-\frac{a^{+}_{in} a^2_{in}}{4S}- \frac{a^{+2}_{in} a_{in}}{4S}\right), \\
  S^{z^{\prime}}_{in}&\approx& -i\sqrt{\frac{S}{2}}\left(a_{in}-a^{+}_{in}-\frac{a^{+}_{in} a^2_{in}}{4S}+\frac{a^{+2}_{in} a_{in}}{4S}\right). \nonumber
\end{eqnarray}
After simple calculations, one obtains that there are no terms in the Hamiltonian which are linear in Bose operators and which contain products of three Bose operators. Terms containing products of two operators of creation and annihilation have the form
\begin{eqnarray}
  \label{H0quant}
	\mathcal{H}_2 &=& S J_0 \sum_{in} \left( 2a^{+}_{in} a_{in} - a^{+}_{in} a_{in+1} - a^{+}_{in} a_{in-1}\right)
  + 2S J_1 \sum_{\langle ij\rangle n} \left( a^{+}_{in} a_{in} - a^{+}_{in} a_{jn}\right),
\end{eqnarray}
where we omit terms of the second order in $D/J_0\ll1$.

\subsection{Perturbation of the magnetic ordering by defects}

Let us discuss a defect bond with DM vector ${\bf D}'=(0,0,D')\ne \bf D$ and $J_0'\ne J_0$ between spins at sites $00$ and $01$ (see Fig.~\ref{Zmagnet}(a)). The following additional terms arise in Hamiltonian \eqref{H0start}:
\begin{eqnarray}
  \label{Vpert}
	\mathcal{V} &=& \mathcal{V}_{dm}+\mathcal{V}_{ex}=-u_{dm} {\bf e}_z \cdot\left[ {\bf S}_{00}\times {\bf S}_{01} \right] - u_{ex} {\bf  S}_{00} \cdot {\bf  S}_{01},\\
	u_{dm} &=& D'-D,\\
	u_{ex} &=& J_0'-J_0.
\end{eqnarray}
One obtains for the perturbation of Hamiltonian \eqref{H0final} from Eq.~\eqref{Vpert} using Eqs.~\eqref{transform} and \eqref{SpinRep}
\begin{eqnarray}
  \label{DMpert}
  \mathcal{V}_{dm} &=& - S u_{dm}\sqrt{\frac{S}{2}} (a^{+}_{01}+a_{01}-a^{+}_{00}-a_{00}),\\
  \label{FMpert}
	\mathcal{V}_{ex} &=&
	S u_{ex}\sqrt{\frac{S}{2}}\frac{D}{J_0} (a^{+}_{01}+a_{01}-a^{+}_{00}-a_{00})
	+
	S u_{ex} (a^{+}_{01} a_{01} + a^{+}_{00} a_{00} - a^{+}_{01} a_{00}- a^{+}_{00} a_{01}) ,
\end{eqnarray}
where we take into account only linear and bilinear terms in Bose-operators which are of the zeroth and of the first orders in DMI (in particular, we put $\cos q=1$ and $\sin q= D/J_0$).

Terms in Eqs.~\eqref{DMpert} and \eqref{FMpert} linear in Bose-operators signify a distortion of the FM ordering around the imperfect bond. To eliminate the linear terms in the Hamiltonian, one has to make the shift
\begin{eqnarray}
  \label{shift}
	a_{in} &=& b_{in}+\rho_{in} e^{i \varphi_{in}}, \\
  a^+_{in} &=& b^+_{in}+\rho_{in} e^{-i \varphi_{in}}, \nonumber
\end{eqnarray}
where $\rho_{in}$ and $\varphi_{in}$ are constants (the ``condensate density'' and the ``phase'', respectively) which describe perturbation of the spin ordering due to the defect. As the easy-plane anisotropy in the Hamiltonian forces all spins to lie within the $xy$ plane and we do not consider AF coupling on the defect bond (i.e., $J_0'>0$), we put $\varphi_{in}=0$ in the following to eliminate the magnetization component perpendicular to the easy axis. As it is seen from Eq.~\eqref{SpinRep} and illustrated by inset in Fig.~\ref{rhoplot}(a), a real $\rho_{in}\ne0$ describes a rotation of the magnetization at site $in$ within $xy$ plane. To restrict ourselves to terms of leading powers in $\rho_{in}$ in subsequent calculations, we assume that $|\rho_{in}|\ll\sqrt{S}$. Then, the rotation angle is equal approximately to $\rho_{in}\sqrt{2/S}$ (because $S_{in}^{y\prime}\approx\sqrt{2S} \rho_{in}$ and $S_{in}^{x\prime}=S-\rho_{in}^2 \approx S$).

Bilinear part of the Hamiltonian \eqref{H0quant} acquires the following form after shift \eqref{shift}:
\begin{eqnarray}
  \mathcal{H}_2 &=& \mathcal{H}^{(0)}_2+\mathcal{H}^{(1)}_2+\mathcal{H}^{(2)}_2,\\
%  \mathcal{H}^{(0)}_2 &=& 2SJ_0 \sum_{in} [ \rho^2_{in} - \rho_{in}\rho_{in+1}]  + 2SJ_1 \sum_{\langle ij\rangle n} [ \rho^2_{in} - \rho_{in}\rho_{jn}],\\
	\label{h1}
  \mathcal{H}^{(1)}_2 &=& SJ_0 \sum_{in}  b^+_{in} \left(2 \rho_{in}-\rho_{in+1}- \rho_{in-1} \right) + 2SJ_1 \sum_{\langle ij\rangle n} b^+_{in}\left(\rho_{in} - \rho_{jn} \right) + {\rm h.c.},
%	\\
%  \mathcal{H}^{(2)}_2&=&S J_0 \sum_{in} ( 2b^{+}_{in} b_{in} -  b^{+}_{in} b_{in+1}- b^{+}_{in} b_{in-1} ) + 2S J_1 \sum_{\langle ij\rangle n} ( b^{+}_{in} b_{in} - b^{+}_{in} b_{jn}),
\end{eqnarray}
where h.c.\ denote the Hermitian conjugated terms, $\mathcal{H}^{(0)}_2$ does not contain Bose operators, and $\mathcal{H}^{(2)}_2$ is obtained from Eq.~\eqref{H0quant} by the replacement of operators $a$ by operators $b$. One has to dispose of linear in $b_{in}$ and $b^+_{in}$ terms in the Hamiltonian by choosing proper $\rho_{in}$. As usual, a minimum of the classical energy (i.e., the part of the Hamiltonian not containing Bose-operators) is realized at those $\rho_{in}$ which cancel the linear terms in the Hamiltonian. Let us find such $\rho_{in}$ in two steps: we consider first $\mathcal{V}_{dm}$ only assuming that $u_{ex}=0$ and then we take into account both $\mathcal{V}_{dm}$ and $\mathcal{V}_{ex}$ given by Eqs.~\eqref{DMpert} and \eqref{FMpert}, respectively.

\subsubsection{Defects in DMI only ($u_{ex}=0$)}
\label{DMonly}

\begin{figure}
  \noindent
  \includegraphics[scale=0.57]{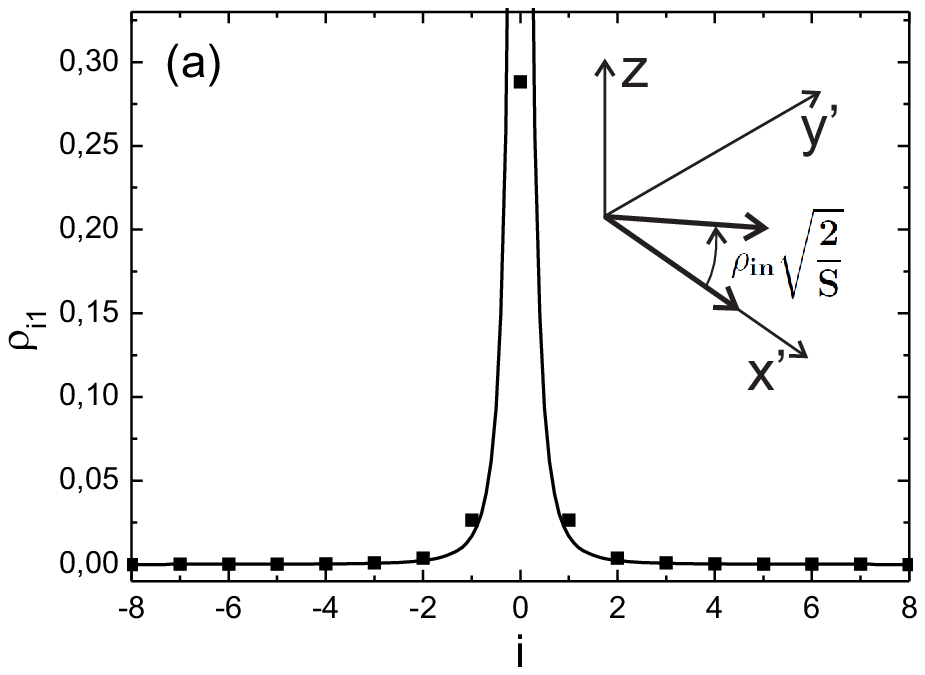}
  \includegraphics[scale=0.33]{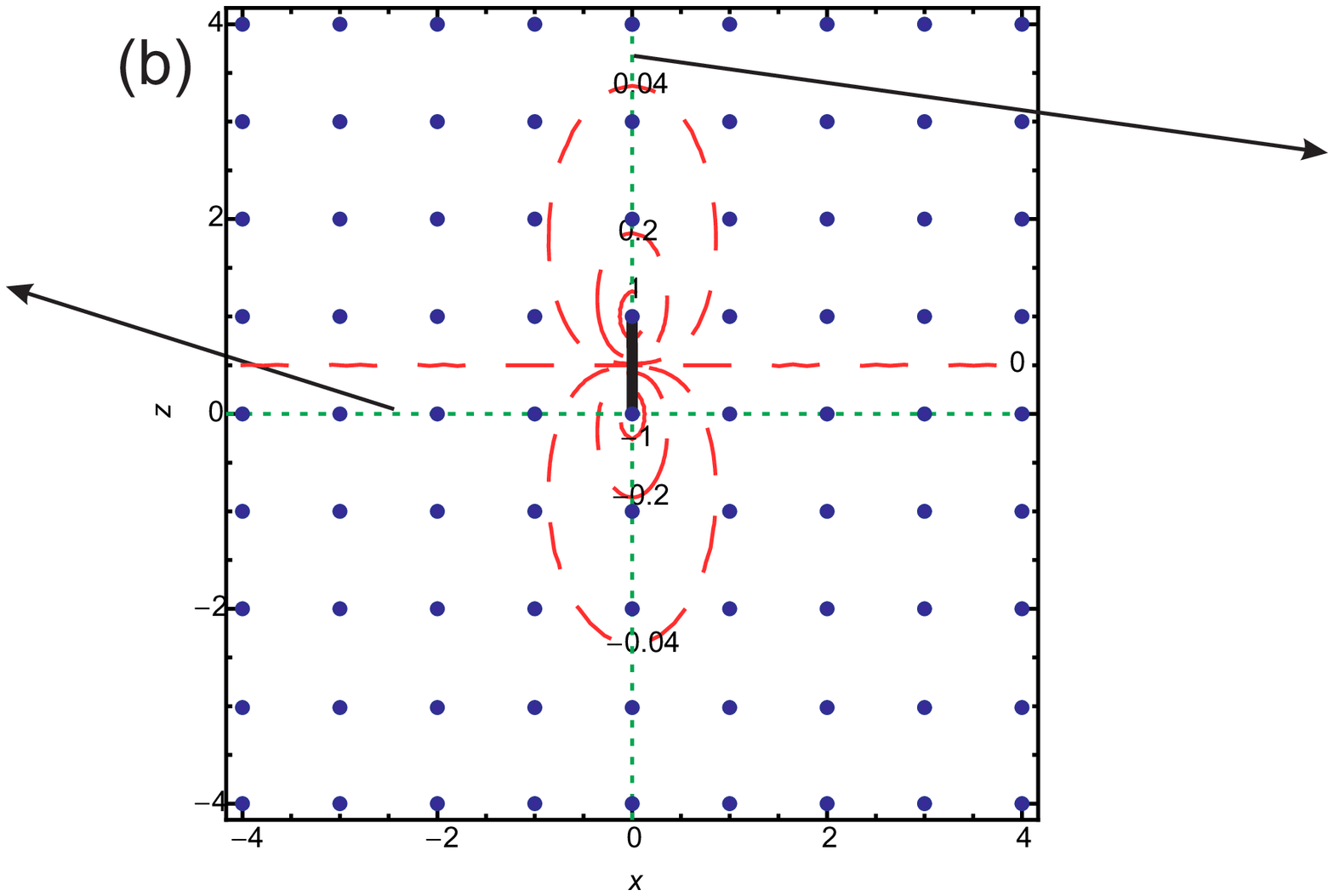}
  \includegraphics[scale=0.57]{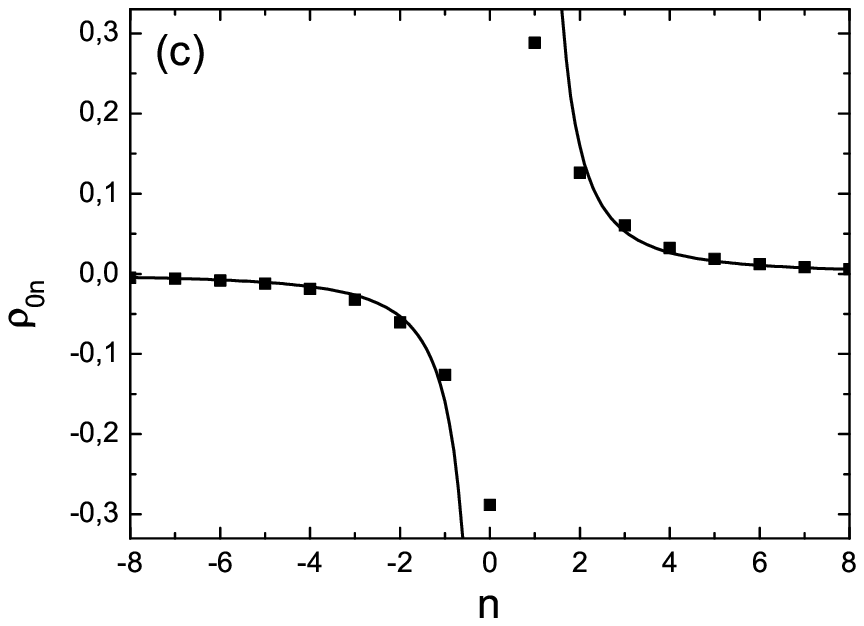}
  \hfil
  \caption{(Color online.)
	  (b) $xz$ plane containing the defect bond which is shown by bold line. Lattice sites are drawn by blue circles. Contour plot is also shown at $J_1=J_0/4$ of the function $\rho({\bf r})$ given by Eq.~\eqref{solsc} and divided by $Q/J_0$.
	 (a) and (c) Plots of $\rho({\bf r})$ along dashed lines depicted in panel (b). Squares represent the result of numerical solution of Eqs.~\eqref{SysDM} for the cluster with $24\times24\times24$ sites. It is seen that the analytical result \eqref{solsc}--\eqref{d} starts working right from sites adjacent to the defect bond.
	Inset in panel (a) illustrates the meaning of the ``condensate density'' $\rho_{in}$ (as well as its counterpart $\rho({\bf r})$ in the continuum limit) in our consideration. Appearance of $\rho_{in}\ne0$ at site $in$ signifies a rotation of the magnetic moment at that site by angle $\rho_{in}\sqrt{2/S}$ in the $xy$ plane.
  \label{rhoplot}}
\end{figure}

We start with the one-impurity problem and then we consider a finite concentration of defects. One has from Eq.~\eqref{DMpert} after shift \eqref{shift}
\begin{eqnarray}
\label{vdm}
  \mathcal{V}_{dm} = u_{dm} S \sqrt{\frac{S}{2}} (2 \rho_{00} - 2 \rho_{01}
  + b^+_{00} +b_{00} -  b^+_{01}- b_{01} ).
\end{eqnarray}
In order linear terms die out in the Hamiltonian, the following equations should hold for every site $in$ which follows from Eqs.~\eqref{h1} and \eqref{vdm}
\begin{eqnarray}
  \label{SysDM}
	J_1\sum_j({\rho}_{in}-{\rho}_{jn}) +J_0(2{\rho}_{in}-{\rho}_{in-1}-{\rho}_{in+1})= -u_{dm}\sqrt{\frac{S}{2}}(\delta_{in,00}-\delta_{in,01}),\qquad\forall in
\end{eqnarray}
where $j$ enumerates nearest neighbors of $i$-th site in $n$-th plane and $\delta$ is the Kronecker delta. It is well known that the second derivative of a function $f(x)$ can be written as
\begin{equation}
  \frac{d^2 f(x)}{d x^2} \approx \frac{f(x+h)+f(x-h)-2f(x)}{h^2}
\end{equation}
with a good precision if $f(x)$ does not change considerably at a distance of $h$. Thus, Eqs.~\eqref{SysDM} can be represented in the differential form in the continuum limit as follows:
\begin{equation}
\label{contlim}
  J_1\left( \frac{\partial^2 \rho({\bf r})}{\partial x^{ 2}} +  \frac{\partial^2 \rho({\bf r})}{\partial y^{ 2}}\right) + J_0  \frac{\partial^2 \rho({\bf r})}{\partial z^2}
	=
	u_{dm}\sqrt{\frac{S}{2}}(\delta({\bf r})-\delta({\bf r}-{\bf r}_0)),
\end{equation}
where $\delta({\bf r})$ and $\delta({\bf r}-{\bf r}_0)$ are delta-functions defining positions of two spins involved in the defect bond and ${\bf r_0}=(0,0,1)$. One expects that the solution of Eq.~\eqref{contlim} describes well the solution of Eqs.~\eqref{SysDM} not very close to the imperfect bond, in which region $\rho({\bf r})$ changes rapidly. After rescaling in $xy$ planes
\begin{equation}
\label{resc}
  \tilde{x}=\sqrt{\frac{J_0}{J_1}}x
	\mbox{ and }
	\tilde{y}=\sqrt{\frac{J_0}{J_1}}y,
\end{equation}
Eq.~\eqref{contlim} turns into the Poisson's equation
\begin{equation}
  \label{DMeq}
  \tilde\Delta {\rho}({\bf r}) = \frac{u_{dm}}{J_1} \sqrt{\frac{S}{2}}(\delta({\bf r})-\delta({\bf r}-{\bf r}_0)),
\end{equation}
where $\tilde\Delta = \partial^2/\partial \tilde x^2 + \partial^2/\partial \tilde y^2 + \partial^2/\partial z^2$. Eq.~\eqref{DMeq} describes electrostatic field of a dipole
\begin{eqnarray}
  \label{solsc}
  \rho({\bf r}) &=& \frac{Q}{4\pi J_1} \left( \frac{1}{|\tilde{\mathbf{r}}-{\mathbf{r}}_0|}-\frac{1}{\tilde{r}}\right),\\	
	\label{q}
	Q &=& u_{dm}\sqrt{\frac{S}{2}},\\
	\label{d}
  {\bf d} &=& \frac{1}{4\pi}\frac{Q}{J_1}{\bf e}_z,
\end{eqnarray}
where $\tilde{\bf r} = ({\tilde x,\tilde y,z})$ and $\bf d$ is the dipole moment. Then, the magnetic ordering distortion produced by one defect bond is long-range: it decays with the distance $r$ as $1/r^2$.  We observe by numerical solution of Eqs.~\eqref{SysDM} that the result \eqref{solsc} starts working well right from sites neighboring to the defect bond in a broad range of parameters (see Fig.~\ref{rhoplot}).

\begin{figure}
  \noindent
  \includegraphics[scale=0.4]{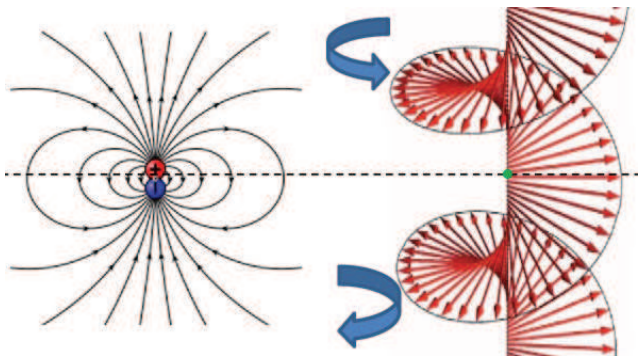}
  \hfil
  \caption{(Color online.) Illustration of the correspondence between the electrostatic picture proposed in the text and the spiral magnet \eqref{H0start}. Horizontal dashed line represents the plane perpendicular to the defect bond and to the dipole moment \eqref{d}. The condensate density $\rho({\bf r})$ given by Eq.~\eqref{solsc} (the field of the dipole) has opposite signs above and below this plane. This signifies that spins lying above and below this plane acquire additional turns in opposite directions which are depicted by blue arrows and which values are governed by $|\rho({\bf r})|$ as it is explained in the text (see also inset in Fig.~\ref{rhoplot}(a)).
  \label{dipole}}
\end{figure}

The correspondence between the electrostatic picture and the spiral magnet \eqref{H0start} is illustrated by Fig.~\ref{dipole}.

Distortion of the FM ordering in the spin system \eqref{H0final} with a finite concentration $c\ll1$ of such randomly distributed defects is described by the electric field from a set of randomly distributed dipoles having the same dipole moment $\bf d$ given by Eq.~\eqref{d}. Averaging over the system volume, one obtains for the ``electric polarization''
\begin{equation}
  {\bf P}=\frac{c}{\tilde{v}_0} {\bf d},
\end{equation}
where $\tilde{v}_0=J_0/J_1$ is the unit cell volume after rescaling \eqref{resc}. The field $\overline{\rho}({\bf r})$ inside the uniformly polarized system is given by the equation
\begin{equation}
\label{unsol}
  \vec\nabla{\overline{\rho}({\bf r})}= 4 \pi {\bf P}
\end{equation}
which has the following explicit form in our case:
\begin{equation}
\label{eq}
  \frac{\partial \overline{\rho}({\bf r})}{\partial z}=\frac{c}{\tilde{v}_0}\frac{u_{dm}}{J_1}\sqrt{\frac{S}{2}}=c\frac{u_{dm}}{J_0}\sqrt{\frac{S}{2}}.
\end{equation}
The solution of Eq.~\eqref{eq} gives an averaging solution of our problem which has the form
\begin{equation}
  \label{Avrho}
	\overline{\rho}({\bf r}) = z c \frac{u_{dm}}{J_0}\sqrt{\frac{S}{2}},
\end{equation}
where we omit a constant corresponding to a rotation of all spins in the system by the same angle. Eq.~\eqref{Avrho} corresponds to the following correction to pitch \eqref{pitch}:
\begin{equation}
  \label{deltaQ1}
  \delta q=\frac{S^{y\prime}_{in}}{S^{x\prime}_{in}}-\frac{S^{y\prime}_{in-1}}{S^{x\prime}_{in-1}}=c \frac{u_{dm}}{J_0}\ll1.
\end{equation}

It should be noted that the requirement $|\rho_{in}|\ll\sqrt{S}$ is essentially important for calculations leading to Eq.~\eqref{Avrho} whereas Eq.~\eqref{Avrho} contradicts it. This discrepancy can be easily removed by applying rotation \eqref{transform} by the pitch $q+\delta q$ rather than by $q$, where $q$ and $\delta q$ are given by Eqs.~\eqref{pitch} and \eqref{deltaQ1}, respectively. Carrying out again the corresponding calculations, we obtain, in particular, Eq.~\eqref{unsol} with $\bf P=0$ which solution is $\overline\rho({\bf r})=0$. Then, we conclude again (not violating the requirement $|\rho_{in}|\ll\sqrt{S}$) that defects lead on average to the correction \eqref{deltaQ1} to the spiral pitch \eqref{pitch}.

After rotation \eqref{transform} by the pitch $q+\delta q$, one obtains for the condensate densities not very close to impurities
\begin{equation}
  \label{newrho}
  \rho_{in} = \sum^{N_d}_{j=1} \frac{\mathbf{d} \cdot \left(\tilde{\mathbf{R}}_{in}-\tilde{\mathbf{R}}_{j}\right)}{\left|\tilde{\mathbf{R}}_{in}-\tilde{\mathbf{R}}_{j}\right|^3} - 4 \pi c \frac{J_1}{J_0}\mathbf{d} \cdot \mathbf{R}_{in},
\end{equation}
where $\bf d$ is given by Eq.~\eqref{d} and $j$ enumerates $N_d$ defect bonds in the system. The first term in Eq.~\eqref{newrho} is the field from all dipoles in the lattice and the second one arises due to the additional turn by $\delta q$. Naturally, averaging of Eq.~\eqref{newrho} over the whole system gives zero.

\subsubsection{Defects both in the exchange interaction and DMI}
\label{general}

Taking into account also the imperfection of the exchange interaction \eqref{FMpert} on the defect bond, we obtain from Eqs.~\eqref{Vpert}, \eqref{DMpert}, and \eqref{FMpert} for the part of $\cal V$ which is linear in Bose operators
\begin{equation}
  \label{gen_pert}
   \mathcal{V}^{(1)} =
	S\left(
	\left(u_{dm}-u_{ex} \frac{D}{J_0} \right) \sqrt{\frac{S}{2}}
 + u_{ex} (\rho_{00}-\rho_{01})
\right)
\left(b^+_{00} +b_{00} -  b^+_{01}- b_{01} \right).
\end{equation}
The counterpart of Eq.~\eqref{SysDM} has the form in this case
\begin{equation}
  \label{SysGen}
  J_1\sum_j({\rho}_{in}-{\rho}_{jn})+ J_0(2{\rho}_{in}-{\rho}_{in-1}-{\rho}_{in+1}) = (\delta_{in,00}-\delta_{in,01})\left[ \left(u_{ex}\frac{D}{J_0}-u_{dm} \right) \sqrt{\frac{S}{2}} + u_{ex} ({\rho}_{01}-{\rho}_{00} )\right],
	\quad \forall in.
\end{equation}
These equations are more complicated than Eqs.~\eqref{SysDM} because one cannot solve them directly in the continuum limit. As it is pointed out above, the solution in the continuum limit does not describe the solution of the initial equations near the defect bond. On the other hand, the solution of Eqs.~\eqref{SysGen} in the continuum limit is essentially determined by condensate densities at sites involved in the defect bond (because the right-hand side depends on $\rho_{01}$ and $\rho_{00}$). Then, we use the following self-consistent scheme to solve Eqs.~\eqref{SysGen}. First, we put
\begin{equation}
  \rho_{01}-\rho_{00}=\alpha
\end{equation}
in the right-hand side of Eqs.~\eqref{SysGen} and treat $\alpha$ as an unknown constant. As a result one returns to the problem considered in the previous subsection which solution is given by Eq.~\eqref{solsc}, where now
\begin{equation}
  Q = \sqrt{\frac{S}{2}}\left( u_{dm} - u_{ex} \frac{D}{J_0}\right) - \alpha u_{ex}.
  \label{Qsc}
\end{equation}
Second, we consider two equations \eqref{SysGen} for $n=0$ and $n=1$
\begin{eqnarray}
  \label{SysSC1}
  4 J_1 (\rho_{00} - \rho_{10}) + J_0 (2\rho_{00}-\rho_{01}-\rho_{0-1}) &=& -u_{ex} (\rho_{00}-\rho_{01})- \sqrt{\frac{S}{2}}\left( u_{dm} - u_{ex} \frac{D}{J_0}\right), \\
  \label{SysSC2}	
	4 J_1 (\rho_{01} - \rho_{11}) +J_0 (2\rho_{01}-\rho_{00}-\rho_{02}) &=& u_{ex} (\rho_{00}-\rho_{01})+ \sqrt{\frac{S}{2}}\left( u_{dm} - u_{ex} \frac{D}{J_0}\right),
\end{eqnarray}
where $\rho_{10}$ and $\rho_{11}$ are condensate densities at sites neighboring to spins involved in the defect bond and lying in planes with $n=0$ and $n=1$, respectively (i.e., we use the system symmetry). One obtains by subtracting Eq.~\eqref{SysSC1} from \eqref{SysSC2}
\begin{equation}
\label{eqal}
  J_1(4 \alpha - 8\rho_{11} ) + J_0 (3 \alpha - 2\rho_{02}) = -2 \alpha u_{ex} + \sqrt{2S}\left( u_{dm} - u_{ex} \frac{D}{J_0}\right),
\end{equation}
where we use that one can turn the coordinate system to fulfill relations $\rho_{02}=-\rho_{0-1}$ and $\rho_{10}=-\rho_{11}$. Using our finding that the result \eqref{solsc} obtained in the continuum limit starts working well right from sites neighboring to the defect bond in a broad range of parameters, we derive $\rho_{02}$ and $\rho_{11}$ from Eqs.~\eqref{solsc} and \eqref{Qsc}. Then, Eq.~\eqref{eqal} turns into an equation for $\alpha$ which solution is given by
\begin{equation}
\label{alpha}
  \alpha=\frac{ \sqrt{\frac{S}{2}}\left( u_{dm} - u_{ex} \frac{D}{J_0}\right)\left[2 + \frac{ J_0}{4 \pi J_1} + \frac{2}{\pi} \left( \sqrt{\frac{J_1}{J_0}} -\sqrt{\frac{J_1}{J_1+J_0}}\right)\right]}{3J_0+4J_1+u_{ex} \left[2 + \frac{ J_0}{4 \pi J_1} + \frac{2 }{\pi} \left( \sqrt{\frac{J_1}{J_0}} -\sqrt{\frac{J_1}{J_1+J_0}}\right)\right]}.
\end{equation}
One has from Eqs.~\eqref{Qsc} and \eqref{alpha}
\begin{equation}
\label{qres}
  Q=\sqrt{\frac{S}{2}}
	\left( u_{dm} - u_{ex} \frac{D}{J_0}\right)
	\frac{3J_0+4J_1}{3J_0+4J_1+u_{ex} \left[2 + \frac{ J_0}{4 \pi J_1} + \frac{2 }{\pi} \left( \sqrt{\frac{J_1}{J_0}} -\sqrt{\frac{J_1}{J_1+J_0}}\right)\right]} .
\end{equation}
It should be noted that Eq.~\eqref{alpha} can give an infinitely large result at $J_1<0.1J_0$ and $-1<u_{ex}<0$. This signifies that more equations \eqref{SysGen} should be considered in addition to Eqs.~\eqref{SysSC1} and \eqref{SysSC2} to find $\rho_{00}$ and $\rho_{01}$. The corresponding analysis is out of the scope of the present paper. Our numerical solutions of Eqs.~\eqref{SysGen} on finite clusters show that Eqs.~\eqref{solsc} and \eqref{qres} work well beyond the region $J_1<0.1J_0\bigcup-1<u_{ex}<0$ not very close to the defect bond.

As it is done above, we derive for the correction to the spiral pitch (cf.\ Eq.~\eqref{deltaQ1})
\begin{equation}
\label{dq}
   \delta q=c \left(u_{dm} - u_{ex} \frac{D}{J_0}\right)
	\frac{3+4J_1/J_0}{3J_0+4J_1+u_{ex} \left[2 + \frac{ J_0}{4 \pi J_1} + \frac{2 }{\pi} \left( \sqrt{\frac{J_1}{J_0}} -\sqrt{\frac{J_1}{J_1+J_0}}\right)\right]}
	\ll1.
\end{equation}
Values of $\rho_{in}$ are determined by Eq.~\eqref{newrho}, where $\bf d$ and $Q$ are given by Eqs.~\eqref{d} and \eqref{qres}, respectively.

It is interesting to note that the influence of imperfections in the DMI and in the exchange interaction on the spiral ordering weaken each other substantially at $u_{dm} \approx u_{ex} {D}/{J_0}$ (see Eqs.~\eqref{qres} and \eqref{dq}).

\subsection{How defect bonds seen in elastic neutron scattering experiments}
\label{ENS1}

\begin{figure}
  \noindent
  \includegraphics[scale=0.8]{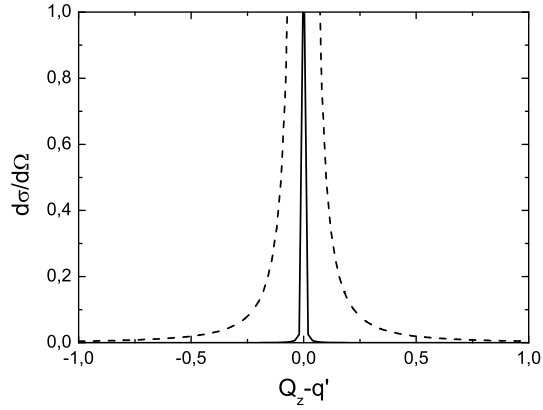}
  \hfil
  \caption{Sketch illustrating Eqs.~\eqref{simpleCS} and \eqref{B20CS} for elastic neutron scattering cross-section at systems with defect bonds. The magnetic Bragg peak at momentum transfer ${\mbox{\boldmath $\cal Q$}}={\bf q}'$ is shown by solid line (${\cal Q}_x={\cal Q}_y=0$). Power-law decaying tails are shown by dashed lines which are given by the second terms in Eqs.~\eqref{simpleCS} and \eqref{B20CS}.}
  \label{neutron}
\end{figure}

The cross-section of elastic neutron scattering is given by \cite{Lowesey}
\begin{equation}
\label{cs}
  \frac{d \sigma}{d \Omega} \propto
	\sum_{in,jm} e^{i {\mbox{\boldmath $\cal Q$}}(\mathbf{R}_{in}-\mathbf{R}_{jm})}
	\sum_{\chi,\eta} (\delta_{\chi\eta}-{\widehat{\cal Q}}^\chi\widehat{\cal Q}^\eta)
	\langle S^\chi_{in}\rangle \langle S^\eta_{jm}\rangle,
\end{equation}
where ${\mbox{\boldmath $\cal Q$}}$ is the momentum transfer, $\widehat{{\mbox{\boldmath $\cal Q$}}}={\mbox{\boldmath $\cal Q$}}/\cal Q$, $\chi,\eta=x,y$, $ \langle \dots \rangle$ denotes an average over quantum and thermal fluctuations,
\begin{eqnarray}
\label{sx}
  \langle S^{x}_{in} \rangle &\approx& S \cos{n q^\prime} - \sqrt{2S} \rho_{in} \sin{n q^\prime} - \rho^2_{in} \cos{n q^\prime}, \\
\label{sy}
  \langle S^{y}_{in} \rangle &\approx& S \sin{n q^\prime} + \sqrt{2S} \rho_{in} \cos{n q^\prime} - \rho^2_{in} \cos{n q^\prime}, \\
  q^\prime&=&q+\delta q,
\end{eqnarray}
$\rho_{in}$ are given by Eq.~\eqref{newrho}, $q$ and $\delta q$ are given by Eqs.~\eqref{pitch} and \eqref{dq}, respectively, and we omit terms of orders higher than the second power of $\rho$. Terms in Eqs.~\eqref{sx} and \eqref{sy} not containing $\rho$ lead to the well known result for a spiral magnet without disorder
\begin{equation}
  \label{PureCS}
	  \left(\frac{d \sigma}{d \Omega}\right)_{\rm Bragg} \propto
  N 2 \pi^3 S^2\left(1+\widehat{{\cal Q}}_z^2\right) \sum_{\mbox{\boldmath $\tau$}}
  \left(\delta({\mbox{\boldmath $\cal Q$}}+\mathbf{q}^\prime-\mbox{\boldmath $\tau$})+\delta({\mbox{\boldmath $\cal Q$}}-\mathbf{q}^\prime-\mbox{\boldmath $\tau$})\right),
\end{equation}
where $N$ is the number of sites in the lattice, ${\bf q}'=q'{\bf e}_z$, $\mbox{\boldmath $\tau$}$ are reciprocal lattice vectors, and delta-functions describe the magnetic Bragg peaks (satellites) at ${\mbox{\boldmath $\cal Q$}}=\pm \mathbf{q}^\prime+\mbox{\boldmath $\tau$}$. Terms in Eq.~\eqref{cs} linear in $\rho$ give zero after averaging over disorder configurations. Most of the second order in $\rho$ terms give either zero or contributions proportional to Eq.~\eqref{PureCS} with a small factor $c Q^2$. The only important quadratic in $\rho$ term has the following structure:
\begin{equation}
  S\left(1 + \widehat{{\cal Q}}_z^2\right)
	\overline{\sum_{in,jm} e^{i {\mbox{\boldmath $\cal Q$}}\left(\mathbf{R}_{in}-\mathbf{R}_{jm}\right)}
	\rho_{in}\rho_{jm} \cos{(m-n)q^\prime}},
\end{equation}
where the line denotes the averaging over disorder configurations. This averaging can be easily carried out using the following expression for the Fourier transform of the field from a single dipole:
\begin{equation}
  \label{DipoleFourier}
  \int d{\bf r} e^{i \mathbf{k}\cdot \mathbf{r}}
	\left( \frac{1}{\left|\tilde{\mathbf{r}}-\tilde{\mathbf{R}}_0-\frac12\mathbf{e}_z\right|}
	-
	\frac{1}{\left|\tilde{\mathbf{r}}-\tilde{\mathbf{R}}_0+\frac12\mathbf{e}_z\right|}\right)
	=
	\frac{4\pi J_1}{\tilde{k}^2J_0}e^{i \tilde{\mathbf{k}}\cdot \tilde{\mathbf{R}}_0} \left( e^{i k_z/2}-e^{-i k_z/2} \right),
\end{equation}
where $\mathbf{R}_0$ specifies the dipole center, $\tilde{\bf k}=(k_x\sqrt{J_1/J_0},k_y\sqrt{J_1/J_0},k_z)$, and ${\bf k}=(k_x,k_y,k_z)$ (cf.\ Eq.~\eqref{resc}). As a result one obtains for the elastic cross section
\begin{equation}
  \label{simpleCS}
  \frac{d \sigma}{d \Omega} \propto
	\left(\frac{d \sigma}{d \Omega}\right)_{\rm Bragg}
	+
	N c S \left( \frac{Q}{J_0}\right)^2 \left(1+\widehat{{\cal Q}}_z^2\right)
	\sum_{\mbox{\boldmath $\tau$}}
	\left(
	\frac{1-\cos{({\cal Q}_z+q^\prime-\tau_z)}}{\left(\widetilde{{\mbox{\boldmath $\cal Q$}}}+\mathbf{q}^\prime-\tilde{{\mbox{\boldmath $\tau$}}}\right)^4}
	+
	\frac{1-\cos{({\cal Q}_z-q^\prime-\tau_z)}}{\left(\widetilde{{\mbox{\boldmath $\cal Q$}}}-\mathbf{q}^\prime-\tilde{{\mbox{\boldmath $\tau$}}}\right)^4}
	\right),
\end{equation}
where the first term is given by Eq.~\eqref{PureCS} and the second one describes the diffuse magnetic scattering due to the disorder. Quite unexpectedly for diffuse scattering caused by disorder, \cite{Lowesey} the second term in Eq.~\eqref{simpleCS} has the power-law singularities at the magnetic Bragg peaks positions (cf.\ Eq.~\eqref{PureCS}). Then, one obtains that Bragg peaks acquire power-law decaying tails (see Fig. \ref{neutron}). This feature is attributed to the long-range character of the perturbation made by defect bonds.

\subsection{Magnon spectrum renormalization in the layered magnet with DMI}
\label{magnon}

In this section, we discuss defects impact on the magnon spectrum in the model \eqref{H0start}. We remind first the well known results for the pure system. 

\subsubsection{Spectrum of the pure system}

One obtains for the bilinear part of the Hamiltonian using Eqs.~\eqref{H0final} and \eqref{SpinRep}
\begin{eqnarray}
\mathcal{H}_2 &=& \sum_{\mathbf{k}} \left[ A_{\bf k} a^+_{\bf k} a_{\bf k} - \frac{B_{\bf k}}{2}\left(a_{\bf k}a_{-\bf k} + a^+_{\bf k} a^+_{-\bf k}\right)\right],\\
  A_{\bf k} &=& 2S(J_0(1-\cos{k_z})+J_1(2-\cos{k_x}-\cos{k_y}))+S \frac{D^2}{2J_0} (2-\cos{k_z}), \\
  B_{\bf k} &=& S\frac{D^2}{2J_0}\cos{k_z}.
\end{eqnarray}
Then, the bare gapless spectrum $\varepsilon^{(0)}_{\bf k}=\sqrt{A^2_{\bf k}-B^2_{\bf k}}$ reads at small $k$ as
\begin{equation}
\label{spec0}
  \varepsilon^{(0)}_{\bf k}=S\sqrt{[J_0k^2_z+J_1(k^2_x+k^2_y)][J_0k^2_z+J_1(k^2_x+k^2_y)+D^2/J_0]}.
\end{equation}
Two regimes can be distinguished
\begin{eqnarray}
\label{reg1}
  \varepsilon^{(0)}_{\bf k} &=& SD\tilde{k}, \qquad \tilde{k} \ll D/J_0, \\
\label{reg2}
  \varepsilon^{(0)}_{\bf k} &=& S J_0 \tilde{k}^2,  \qquad D/J_0 \ll \tilde{k} \ll 1,
\end{eqnarray}
where $\tilde{k}=\sqrt{k^2_z+(k^2_x+k^2_y)J_1/J_0}$.

\subsubsection{Spectrum corrections}

\begin{figure}
  \noindent
  \includegraphics[scale=0.8]{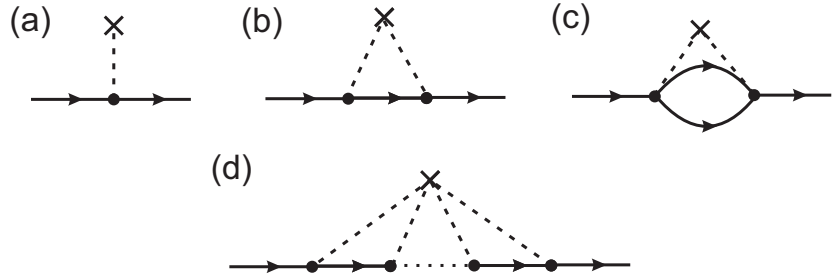}
  \hfil
  \caption{Diagrams giving leading corrections to the magnon spectrum in the first order in the defects concentration.}
  \label{diagrams}
\end{figure}

We imply first that only DMI is changed at imperfect bonds. One has to take into account diagrams shown in Fig.~\ref{diagrams} to find the spectrum corrections. Calculations are simplified by the fact that vertexes in all of the diagrams are proportional to $u_{dm}$ which is much smaller than exchange constants. Some details of the cumbersome diagram analysis can be found in Appendix~\ref{append}, where the following expression for the magnon energy is obtained:
\begin{equation}
\label{speccorr}
  \delta \varepsilon_{\bf k} =
	S c q u_{dm} (2-\cos k_z) + c \frac{u^2_{dm}}{J_0} (I_{1\bf k}+I_{2\bf k}+I_{3\bf k}),
\end{equation}
where $I_{1\bf k}$, $I_{2\bf k}$, and $I_{3\bf k}$ are smooth functions of $\bf k$ which are of the order of unity and which are given by Eqs.~\eqref{i1}, \eqref{i2}, and \eqref{i3}, respectively.

The magnon damping is given by the following term which stems from the diagram shown in Fig.~\ref{diagrams}(b):
\begin{equation}
\label{gamcorr}
  \gamma_{\bf k} = c \frac{k^3}{\varepsilon_{\bf k}}\frac{(Su_{dm}D)^2}{J_0 J_1} \frac{t}{2\pi},
\end{equation}
where $t=1$ and $1/2$ for, respectively, $\tilde{k} \ll D/J_0$ and $\tilde{k} \gg D/J_0$ (see Eqs.~\eqref{spec0}--\eqref{reg2}).

It is seen from Eqs.~\eqref{gamcorr} and \eqref{spec0}--\eqref{reg2} that the magnon damping is much smaller than the bare spectrum at all momenta. In contrast, the correction to the magnon energy \eqref{speccorr} (that is finite at $k=0$) becomes much larger than the bare gapless spectrum at sufficiently small $\bf k$. It implies that our results obtained in the first order in $c$ are inapplicable and further analysis is required for particularly small $\bf k$ that is out of the scope of the present paper. Besides, this effect can be screened in real materials by a small gap in the spectrum originating from, e.g., a small anisotropic interaction.

Let us take into account also the defect in the exchange interaction on imperfect bonds. In contrast to the defect in DMI, one cannot assume in general that $|u_{fm}|\ll J_0,J_1$. Then, one has to sum an infinite set of diagrams of the type shown in Fig.~\ref{diagrams}(d) to find spectrum corrections in the first order in $c$. As a result of tedious calculations some details of which can be found in Appendix~\ref{append}, one leads to the following counterparts of Eqs.~\eqref{speccorr} and \eqref{gamcorr}:
\begin{eqnarray}
\label{speccorrgen}
  \delta \varepsilon_{\bf k} &=&
	S c q \left(u_{dm} - \frac{q u_{ex}}{2}\right) (2-\cos k_z) + c \frac{(u_{dm} - q u_{ex})^2}{J_0} I_{1\bf k}+ c \frac{2 Q^2}{S J_0} I_{2\bf k}+c \sqrt{\frac{2}{S}}\frac{(u_{dm} - q u_{ex}) Q}{J_0}I_{3\bf k},\\
\label{gamcorrgen}
  \gamma_{\bf k} &=&
	c \frac{k^3}{\varepsilon_{\bf k}}S^2\left(u_{dm} - \frac{q u_{ex}}{2}\right)^2\frac{D^2}{J_0 J_1} \frac{t}{2\pi},
\end{eqnarray}
where $Q$ is given by Eq.~\eqref{qres}.

\subsection{Some other layered helimagnets with DMI}
\label{otherlay}

In this subsection, we discuss briefly some other models of layered helimagnets with defect bonds to which our theory is applicable after some modifications. The first model differs from that discussed above by the sign of $J_1$ (i.e., by replacement in Eq.~\eqref{H0start} of $J_1>0$ by $-J_1<0$) that results in AF $xy$ planes. It is convenient to rewrite transformation \eqref{transform} in the following more general form:
\begin{eqnarray}
   \label{sss}
	S^{x}_{in} &=& S^{x^\prime}_{in} \cos{({\bf k}_0 {\bf R}_{in})} - S^{y^\prime}_{in} \sin{({\bf k}_0 {\bf R}_{in})},\nonumber\\
  S^{y}_{in} &=& S^{x^\prime}_{in} \sin{({\bf k}_0 {\bf R}_{in})} + S^{y^\prime}_{in} \cos{({\bf k}_0 {\bf R}_{in})},\\
	S^{z}_{in} &=& S^{z^\prime}_{in},\nonumber
\end{eqnarray}
where ${\bf k}_0=(\pi,\pi,q)$ is the vector of magnetic structure. In pure system, the spiral propagates along $z$ axis and $q$ is given by Eq.~\eqref{pitch}. The operator of perturbation \eqref{gen_pert} and the system of equations for $\rho_{in}$ \eqref{SysGen}, which determine the spin texture around defect, remain the same. Then, Eq.~\eqref{dq} for the corrections to $q$ does not change either.

Let us assume that both $J_0$ and $J_1$ are antiferromagnetic (i.e., we replace in Eq.~\eqref{H0start} $J_0>0$ and $J_1>0$ by $-J_0<0$ and $-J_1<0$, respectively). In this case, ${\bf k}_0=(\pi,\pi,\pi-q)$ in Eq.~\eqref{sss}. The operator of perturbation \eqref{gen_pert} from a single imperfect bond changes its sign that leads to the dipole with opposite dipolar momentum. Then, the vector of magnetic structure acquires the form $(\pi,\pi,\pi-q-\delta q)$, where $\delta q$ is given by Eq.~\eqref{dq}.

The third model contains antiferromagnetic triangular $xy$ planes (i.e., we replace $J_1>0$ by $-J_1<0$ in Eq.~\eqref{H0start} and assume that $xy$ planes are triangular). This model is relevant to $\rm CsCuCl_3$ (see, e.g., Refs.~\cite{nikuni1993,TAF} and references therein). In pure system, $120^\circ$ spin ordering is realized in each $xy$ plane and a spiral ordering arises along $z$ axis. Then, the vector of the magnetic structure ${\bf k}_0$ can be equal either to $(0,\frac{4\pi}{3},q)$ or to $(0,-\frac{4\pi}{3},q)$ (we assume for simplicity that distances between all nearest neighbor spins are equal to unity) which describe $120^\circ$ spin structures with different arrangements of chiralities of triangles in $xy$ planes (see, e.g., Ref.~\cite{TAF}). One obtains the same operator of perturbation \eqref{gen_pert}. The system of linear equations has the form \eqref{SysGen}, where $J_1$ should be replaced by $J_1/2$ and one has to take into account that there are six nearest neighbor spins in $xy$ plane. In the continuum limit, we obtain Eq.~\eqref{contlim} in which $J_1$ should be replaced by $3 J_1/4$. Counterparts of Eqs.~\eqref{qres} and \eqref{dq} have the form
\begin{eqnarray}
  Q &=& 3 \sqrt{\frac{S}{2}}
	\left( u_{dm} - u_{ex} \frac{D}{J_0}\right)
	\frac{J_0+J_1}{3J_0+3J_1+u_{ex} \left[2 + \frac{ J_0}{3 \pi J_1} + \frac{2 }{\pi} \left( \sqrt{\frac{3 J_1}{4 J_0}} -\sqrt{\frac{3J_1}{3J_1+4J_0}}\right)\right]} ,\\
   \delta q &=& c \left(u_{dm} - u_{ex} \frac{D}{J_0}\right)
	\frac{1+J_1/J_0}{J_0+J_1+\frac{u_{ex}}{3} \left[2 + \frac{ J_0}{3 \pi J_1} + \frac{2 }{\pi} \left( \sqrt{\frac{3 J_1}{4 J_0}} -\sqrt{\frac{3J_1}{3J_1+4J_0}}\right)\right]}
	\ll1.
\end{eqnarray}
The vector of magnetic structure has the form $(0,\pm\frac{4\pi}{3},q+\delta q)$. Changing of the sign of $J_0$ in this model leads to the replacement of $Q$ by $-Q$ and to the vector of magnetic structure $(0,\pm\frac{4\pi}{3},\pi-q-\delta q)$.

Results for the spectrum corrections in these systems are not simple modifications of those obtained above for the ferromagnetic exchange because all these models have different bare spectra. Corresponding calculations are out of the scope of the present paper.

\section{Cubic B20 magnets}
\label{B20magnets}

\subsection{Pure cubic B20 magnets}

Our consideration of cubic B20 magnets is based on Refs.~\cite{Bak,maleyev} which are devoted to pure systems. We present in this subsection the well-known results which are important for further analysis of disordered systems. For discussion of low-energy dynamics, the following Hamiltonian is proposed which contains the exchange coupling $\mathcal{H}_{ex}$, DM term $\mathcal{H}_{dm}$, and small anisotropic exchange interaction (AEI) $\mathcal{H}_{ae}$:
\begin{eqnarray}
 \label{HMnSi}
  \mathcal{H}_0 &=& \mathcal{H}_{ex}+\mathcal{H}_{dm}+\mathcal{H}_{ae}, \\
  \mathcal{H}_{ex} &=& -\frac12 \sum J_{{\bf R}{\bf R}^\prime} \mathbf{S}_{\bf R} \cdot \mathbf{S}_{{\bf R}^\prime}, \\
  \mathcal{H}_{dm}&=& - \frac{1}{2}  \sum {\bf D}_{{\bf R}{\bf R}^\prime} \cdot [\mathbf{S}_{\bf R} \times \mathbf{S}_{{\bf R}^\prime}], \\
  \mathcal{H}_{ae} &=& \frac12\sum_\nu F_{{\bf R}{\bf R}^\prime} (\partial_\nu S^\nu_{\bf R})(\partial_\nu S^\nu_{{\bf R}^\prime}),
\end{eqnarray}
where summations on $\bf R$ and $\bf R'$ are taken over all sites of a simple cubic lattice and $\nu=x,y,z$. As it is frequently done in theoretical considerations, we take the cubic lattice structure rather than the full B20 structure mainly for technical simplicity. Besides, a little is known now about interaction between four magnetic ions in the cubic unit cell of the widely discussed itinerant materials having B20 structure. It is assumed that all interactions in Eq.~\eqref{HMnSi} act between nearest neighbor spins: $J_{{\bf R}{\bf R}^\prime}=J$, $D_{{\bf R}{\bf R}^\prime}=D$, ${\bf D}_{{\bf R}{\bf R}^\prime} || ({\bf R}-{\bf R}')$, and $F_{{\bf R}{\bf R}^\prime}=F$. We imply below that $J \gg D \gg F$ and put the lattice constant to be equal to unity. The following local orthogonal coordinate frame is defined at each site:
\begin{eqnarray}
    \mbox{\boldmath $\zeta$}_{\bf R} &=& {\mbox{\boldmath $\mathfrak a$}} \cos (\mathbf{q}\cdot\mathbf{R}) + {\mbox{\boldmath $\mathfrak b$}} \sin (\mathbf{q}\cdot\mathbf{R}), \\
    \mbox{\boldmath $\eta$}_{\bf R} &=& {\mbox{\boldmath $\mathfrak b$}} \cos (\mathbf{q}\cdot\mathbf{R}) - {\mbox{\boldmath $\mathfrak a$}} \sin (\mathbf{q}\cdot\mathbf{R}), \\
    \mbox{\boldmath $\xi$}_{\bf R} &=& {\mbox{\boldmath $\mathfrak c$}},
\end{eqnarray}
where ${\mbox{\boldmath $\mathfrak a$}} \times {\mbox{\boldmath $\mathfrak b$}}= {\mbox{\boldmath $\mathfrak c$}}$. Spins are represented in the local coordinate system as $\mathbf{S}_{\bf R}=S^\zeta_{\bf R}  \mbox{\boldmath $\zeta$}_{\bf R}+S^\eta_{\bf R} \mbox{\boldmath $\eta$}_{\bf R} + S^\xi_{\bf R} \mbox{\boldmath $\xi$}_{\bf R}$. We use the Holstein-Primakoff representations \eqref{SpinRep} for spins components $S^{\zeta,\eta,\xi}_{\bf R}$ with the following axes correspondence: $x^\prime \leftrightarrow \zeta$, $y^\prime \leftrightarrow \eta$, and $ z \leftrightarrow \xi $.

The ground state energy per unit cell has the form at $q\ll 1$
\begin{equation}
  E_{cl}=-JS^2\left(3-\frac{q^2}{2}\right)-  DS^2({\bf q}\cdot {\mbox{\boldmath $\mathfrak c$}}) + \frac 32 S^2F I,
  \label{E_cl2}
\end{equation}
where $I=\sum_\nu q^2_\nu (\mathfrak a^2_\nu+\mathfrak b^2_\nu)$. Obviously, $E_{cl}$ is minimal if ${\bf q}\| {\mbox{\boldmath $\mathfrak c$}}$, i.e., spins rotate in the plane perpendicular to $\bf q$. The direction of $\bf q$ relative to the lattice is determined by the last term in Eq.~\eqref{E_cl2}. For $F>0$, $\bf q$ should be directed along the cube edge to minimize the cubic invariant $I$. If $F<0$, one infers that the helix vector is oriented along one of the cubic space diagonals and $I=2q^2/3$. In both cases one has
\begin{equation}
  \mathbf{q}=\frac{D}{J}{\mbox{\boldmath $\mathfrak c$}}.
\end{equation}
The main role of AEI is to determine the $\bf q$ direction and it can be omitted in other calculations due to its smallness. As $F<0$ in many B20 magnets including MnSi, we discuss this case below. Henceforth, $\mbox{\boldmath $\mathfrak c$}$ is directed along one of the cubic space diagonals.

The bosonic analog of spin Hamiltonian \eqref{HMnSi} has no terms linear in Bose-operators and one has for the bilinear terms
\begin{eqnarray}
  \label{B20Hex}
\mathcal{H}^{(2)}_{ex} &=& JS\sum_{\mathbf{R},\nu}
\left[(a^+_{\bf R} a_{\bf R} + a^+_{{\bf R}+{\bf e}_\nu} a_{{\bf R}+{\bf e}_\nu})\left(1-\frac{q^2_\nu}{2}\right)+
    (a_{\bf R} a_{{\bf R}+{\bf e}_\nu}+a^+_{\bf R} a^+_{{\bf R}+{\bf e}_\nu}) \frac{q^2_\nu}{4} \right.\nonumber\\
		&&{}-	\left.
    (a^+_{\bf R} a_{{\bf R}+{\bf e}_\nu}+a_{\bf R} a^+_{{\bf R}+{\bf e}_\nu})\left(1-\frac{q^2_\nu}{4}\right) \right],\\
		  \label{B20Hdm}
\mathcal{H}^{(2)}_{dm} &=&\frac{1}{3} DSq\sum_{\mathbf{R},\nu} \Bigl[(a^+_{\bf R} a_{\bf R} + a^+_{{\bf R}+{\bf e}_\nu} a_{{\bf R}+{\bf e}_\nu})-
    \frac{1}{2}(a_{\bf R} a_{{\bf R}+{\bf e}_\nu}+a^+_{\bf R} a^+_{{\bf R}+{\bf e}_\nu}+a^+_{\bf R} a_{{\bf R}+{\bf e}_\nu}+a_{\bf R} a^+_{{\bf R}+{\bf e}_\nu}) \Bigr],
\end{eqnarray}
where $\nu=x,y,z$ and ${\bf e}_\nu$ are basis vectors of the cubic lattice.
%We neglect umklapp terms here.

\subsection{Perturbation of the magnetic ordering by defects}

Let us consider an imperfect bond between sites $\mathbf{R}_0=(0,0,0)$ and $\mathbf{R}_1=\mathbf{e}_z=(0,0,1)$. The perturbation in the Hamiltonian has the following form:
\begin{eqnarray}
  \mathcal{V}&=&\mathcal{V}_{dm}+\mathcal{V}_{ex}
	= -
u_{dm} \left({\bf e}_z \cdot [\mathbf{S}_{{\bf R}_0} \times \mathbf{S}_{{\bf R}_1}]\right)  -u_{ex} \mathbf{S}_{{\bf R}_0} \cdot \mathbf{S}_{{\bf R}_1}.
 \end{eqnarray}
Omitting terms containing products of more than two Bose operators, one derive for $\mathcal{V}_{dm}$ and $\mathcal{V}_{ex}$
\begin{eqnarray}
\mathcal{V}_{dm} &=& \frac{1}{3}S u_{dm} q \Bigl[ a^+_0 a_0 + a^+_1 a_1 - \frac{1}{2}(a_0 a_1 + a^+_0 a^+_1+a^+_0 a_1 +a^+_1 a_0)\Bigr]+ \frac{S u_{dm}}{\sqrt{3}}  \sqrt{\frac{S}{2}}(a_0+a^+_0-a_1-a^+_1),  \\
 \label{B20FMpert}
\mathcal{V}_{ex} &=& S u_{ex} \Bigl[ (a^+_0 a_0 + a^+_1 a_1) \left(1-\frac{q^2_z}{2}\right) + (a_0 a_1 + a^+_0 a^+_1)\frac{q^2_z}{4}
  -(a^+_0 a_1 + a^+_1 a_0)\left(1-\frac{q^2_z}{4}\right) + \\
	&&{}+ q_z\sqrt{\frac{S}{2}}(a_1+a^+_1-a_0-a^+_0)\Bigr], \nonumber
\end{eqnarray}
where indexes 0 and 1 stand for $\mathbf{R}_0$ and $\mathbf{R}_1$, respectively. To dispose of terms in the Hamiltonian linear in Bose-operators, we make the shift similar to \eqref{shift} which we write in the form
\begin{equation}
  a_{\bf R} = b_{\bf R} + \tilde\rho_{\bf R} = b_{\bf R} + \rho^\prime_{\bf R}+i\rho^{\prime\prime}_{\bf R},
\end{equation}
where $ \rho^\prime_{\bf R}$ and $\rho^{\prime\prime}_{\bf R}$ are real. Simple but tedious calculations show that the following conditions should hold in order terms in the Hamiltonian vanish which are linear in operators $b_{\bf R}$ and $b^+_{\bf R}$:
\begin{eqnarray}
\label{sysb20}
&&\sum_\nu \left[ J \left(2\tilde{\rho}_{\bf R} - \tilde{\rho}_{{\bf R}-{\bf e}_\nu} - \tilde{\rho}_{{\bf R}+{\bf e}_\nu}  - q^2_\nu \tilde\rho_{\bf R} + \frac{q^2_\nu}{2} (\rho^\prime_{{\bf R}+{\bf e}_\nu}+\rho^\prime_{{\bf R}-{\bf e}_\nu})\right) + \frac{1}{3} D q ( 2\tilde{\rho}_{\bf R} -\rho^\prime_{{\bf R}+{\bf e}_\nu}-\rho^\prime_{{\bf R}-{\bf e}_\nu} )\right]
\nonumber\\
&&{}=  \sqrt{\frac{S}{2}} (q_z u_{ex} -  u_{dm}/\sqrt{3}) (\delta_{{\bf R},{\bf R}_0}-\delta_{{\bf R},{\bf R}_1}) - \left[ u_{ex} (\tilde{\rho}_{{\bf R}_0}-\tilde{\rho}_{{\bf R}_1} + \frac{q^2_z}{2}(\rho^\prime_{{\bf R}_1} -\tilde{\rho}_{{\bf R}_0} )) + \frac{1}{3}u_{dm} q (\tilde{\rho}_{{\bf R}_0} - \rho^\prime_{{\bf R}_1}) \right] \delta_{{\bf R},{\bf R}_0}
\nonumber\\
&&{}- \left[ u_{ex} (\tilde{\rho}_{{\bf R}_1}-\tilde{\rho}_{{\bf R}_0} + \frac{q^2_z}{2}(\rho^\prime_{{\bf R}_0} -\tilde{\rho}_{{\bf R}_1} )) + \frac{1}{3} u_{dm} q (\tilde{\rho}_{{\bf R}_1} - \rho^\prime_{{\bf R}_0}) \right] \delta_{{\bf R},{\bf R}_1}, \qquad \forall {\bf R}.
\end{eqnarray}
Imaginary parts of equations \eqref{sysb20} form a linear homogeneous system of equations for $\rho^{\prime\prime}_{\bf R}$ which gives $\rho^{\prime\prime}_{\bf R}=0$. Real parts of Eqs.~\eqref{sysb20} give equations for $\rho^{\prime}_{\bf R}$ which have the form similar to that of Eq.~\eqref{SysGen}
\begin{eqnarray}
\label{sysb202}
&&\sum_\nu \left( J \left( 1-\frac{q^2_\nu}{2}\right) + \frac{1}{3} D q \right)(2\rho^{\prime}_{\bf R} -\rho^{\prime}_{{\bf R}-{\bf e}_\nu} - \rho^{\prime}_{{\bf R}+{\bf e}_\nu} )
\nonumber\\
&&{}= (\delta_{{\bf R},{\bf R}_0}-\delta_{{\bf R},{\bf R}_1})
     \left[ \sqrt{\frac{S}{2}} (q_z u_{ex} -  u_{dm}/\sqrt{3}) + \left(u_{ex} \left( 1-\frac{q^2_z}{2}\right) + \frac{1}{3} u_{dm} q\right)(\rho^{\prime}_{{\bf R}_0}-\rho^{\prime}_{{\bf R}_1})\right], \qquad \forall {\bf R}.
\end{eqnarray}
Eqs.~\eqref{sysb202} can be solved as Eqs.~\eqref{SysGen} with the result (cf.\ Eq.~\eqref{solsc})
\begin{equation}
  \label{solscb20}
  \rho'({\bf R}) = \sqrt{\frac{S}{2}} \frac{Q}{4\pi} \left( \frac{1}{|\mathbf{R}-\mathbf{R}_1|}-\frac{1}{{|\mathbf{R}-\mathbf{R}_0|}}\right).	
\end{equation}
Then, $\rho^\prime_{\bf R}$ is given by the field of a dipole which momentum has the form
\begin{eqnarray}
{\bf d} &=& {\bf e}_z\sqrt{\frac{S}{2}}\frac{Q}{4 \pi}, \\
\label{b20q}
  Q &=& \frac{7 \alpha}{7+ t \beta},
\end{eqnarray}
where $t= 2-(\sqrt2-9/4)/\pi\approx2.27$ and
\begin{eqnarray}
\label{alp}
  \alpha &=& \frac{ u_{dm} - q u_{ex}}{\sqrt{3}J(1 + q^2/6)}, \\
	\label{beta}
  \beta &=& \frac{u_{ex} + u_{dm} q/3 - u_{ex}q^2/6}{J}.
\end{eqnarray}

\begin{figure}
  \noindent
  \includegraphics[scale=0.8]{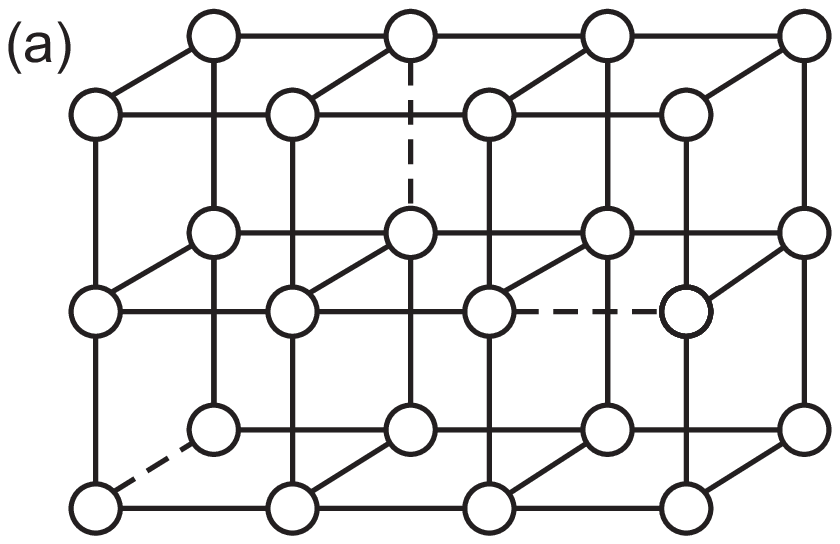}
  \includegraphics[scale=0.8]{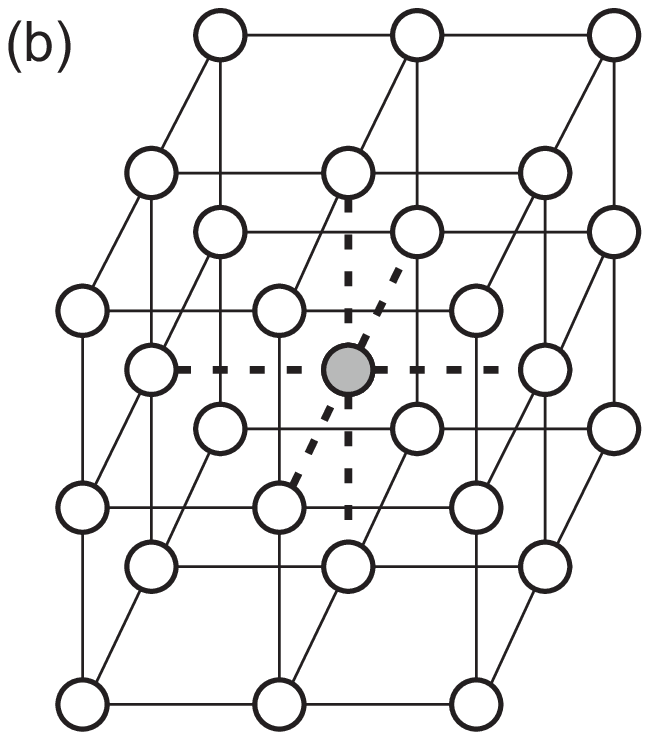}
  \hfil
  \caption{Defects in cubic magnets considered in the text. (a) Imperfect bonds (marked by dashed lines) which can be oriented along each cubic axis with equal probability. (b) Defect with six imperfect bonds which can be relevant to mixed compounds of the type Mn$_{1-x}$Fe$_x$Ge at $x\ll1$ or $x\approx1$.}
  \label{b20}
\end{figure}

Let us turn to the system with a finite concentration $c\ll1$ of such defects. We assume that randomly distributed imperfect bonds orient randomly along three cubic axes as it is illustrated by Fig.~\ref{b20}(a). In this case, a finite correction to the spiral pitch arises. In terms of the electrostatic analogy, the system ``polarization''
$
  \mathbf{P} = c d /\sqrt{3} {\mbox{\boldmath $\mathfrak c$}}
$
arises that is directed along the cubic space diagonal. Correspondingly, the correction to the spiral pitch $q$ has the form
\begin{equation}
\label{dqb20}
  \delta q = c \frac{Q}{\sqrt{3}}.
\end{equation}
A substantial reduction should be pointed out of defects impact on the system properties at $u_{dm} \approx u_{ex} q$ that follows from Eqs.~\eqref{b20q} and \eqref{alp}.

\begin{figure}
  \noindent
  \includegraphics[scale=0.6]{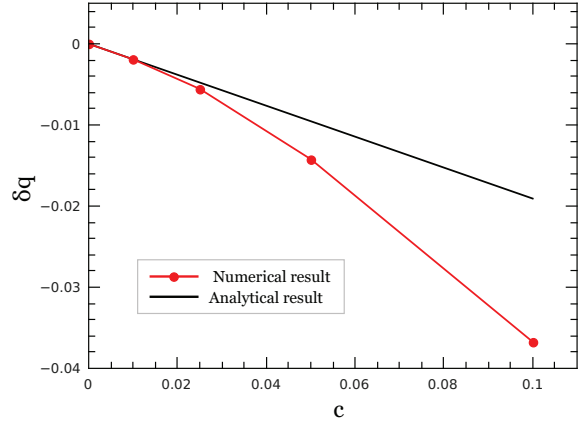}
  \hfil
  \caption{(Color online.) Correction $\delta q$ to the spiral vector as a function of defect concentration $c$ for $J=1$, $D=0.3$, $u_{ex}=-0.2$, and $u_{dm}=-0.6$. Analytical result is given by Eqs.~\eqref{b20q}--\eqref{dqb20}. Numerical result is obtained as it is discussed in the text.}
  \label{num}
\end{figure}

To verify Eqs.~\eqref{b20q}--\eqref{dqb20}, we perform numerical calculations for a set of model parameters. We minimize the classical energy of clusters with open boundary conditions containing up to $100^3$ sites in the following way. Starting from a trial configuration, we arrange all magnetic moments along their current molecular fields.
After performing this procedure many times ($\sim10^6\div10^7$), the system stabilizes and we take the Fourier transformation of the final configuration (ignoring spins near the cluster boundary) which has a peak at the spiral vector $\bf q$ (for the given disorder realization).
Averaging over $10\div20$ disorder realizations, one obtains the spiral vector. Representative results of such calculations are shown in Fig.~\ref{num}. It is seen that the agreement is excellent at $c<0.03$ of numerical findings with Eqs.~\eqref{b20q}--\eqref{dqb20}.

Another type of defects in B20 magnets which we consider is presented in Fig.~\ref{b20}(b). It looks more natural for mixed compounds Mn$_{1-x}$Fe$_x$Ge considered recently experimentally: one expects that substitution of one magnetic atom by another in a unit cell of itinerant material changes couplings of this unit cell with all its neighbors. The system of equations \eqref{sysb20} describing distortion of the spiral ordering caused by one imperfect bond is linear. Then, the result for the considered type of defect is a linear combination of solutions for six defect bonds shown in Fig.~\ref{b20}(b). As a consequence, the ``polarization'' and the correction to the spiral pitch are six times as large as those for one imperfect bond:
$
  \mathbf{P} = 2\sqrt{3} c d  {\mbox{\boldmath $\mathfrak c$}}
$ and
\begin{equation}
\label{dqb20_2}
  \delta q =  2\sqrt{3} c Q
\end{equation}
(cf.\ Eq.~\eqref{dqb20}).

\subsection{Elastic neutron scattering}

For the model of imperfect bonds shown in Fig.~\ref{b20}(a), the main difference from layered magnets discussed above is that there are dipoles with momenta directed along three cubic axes. The concentration of dipoles directed along each cubic axis is $c/3$. Taking this into account, we get the following expression for the elastic neutron scattering cross-section after tedious calculations (cf.\ Eq.~\eqref{simpleCS}):
\begin{equation}
  \label{B20CS}
  \frac{d \sigma}{d \Omega} \propto
	\left(\frac{d \sigma}{d \Omega}\right)_{\rm Bragg}
	+
	N \frac{c}{3} S \left( \frac{Q}{J_0}\right)^2 \left(1+\widehat{{\cal Q}}_c^2\right)
	\sum_{\mbox{\boldmath $\tau$}} \sum_{\nu=x,y,z}
	\left(
	\frac{1-\cos{({\cal Q}_\nu+q^\prime_\nu-\tau_\nu)}}{\left(\widetilde{{\mbox{\boldmath $\cal Q$}}}+\mathbf{q}^\prime-\tilde{{\mbox{\boldmath $\tau$}}}\right)^4}
	+
	\frac{1-\cos{({\cal Q}_\nu-q^\prime_\nu-\tau_\nu)}}{\left(\widetilde{{\mbox{\boldmath $\cal Q$}}}-\mathbf{q}^\prime-\tilde{{\mbox{\boldmath $\tau$}}}\right)^4}
	\right),
\end{equation}
where the first term is given by Eq.~\eqref{PureCS}. Then, Bragg peaks in B20 magnets acquire power-law decaying tails (see Fig.~\ref{neutron}). The last term in Eq.~\eqref{B20CS} should be multiplied by 6 in the case of defects shown in Fig.~\ref{b20}(b).

\subsection{Magnon spectrum renormalization}

It is well known \cite{maleyev,Bak} that the bare magnon spectrum obtained from Eqs.~\eqref{B20Hex} and \eqref{B20Hdm} has the form
\begin{equation}
\label{bareb20}
\begin{aligned}
  \varepsilon^{(0)}_{\bf k} &= S J q k, \qquad k \ll D/J, \\
  \varepsilon^{(0)}_{\bf k} &= S J k^2, \qquad D/J\ll k \ll 1.
\end{aligned}
\end{equation}
There is also a small gap in the spectrum which can be a result of magnon-magnon and magneto-elastic interactions. \cite{maleyev2} This gap is important for interpretation of some experimental data obtained in B20 magnets. \cite{maleyev,maleyev2,grig2009}

Carrying out calculations similar to those for layered spiral magnets, we obtain for corrections to the spectrum
\begin{eqnarray}
\label{deb20}
  \delta \varepsilon_{\bf k} &=& \sum_\nu \frac{S c q}{9}\left( u_{dm}- \frac{q u_{ex}}{2}\right)(2- \cos k_\nu) + c \frac{(u_{dm}-q u_{ex})^2}{3 J} I_{1\bf k} + c \frac{Q^2}{J} I_{2\bf k} + c \frac{(u_{dm} - q u_{ex}) Q}{\sqrt{3} J} I_{3\bf k},\\
	\label{gamb20}
  \gamma_{\bf k} &=&
	c \frac{k^3}{\varepsilon_{\bf k}}S^2\left(u_{dm} - \frac{q u_{ex}}{2}\right)^2 q^2 \frac{t}{18\pi},
\end{eqnarray}
where $\nu=x,y,z$ and $I_{1\bf k}$, $I_{2\bf k}$, and $I_{3\bf k}$ are values of the order of unity which are smooth functions of ${\bf k}$ having the following form:
\begin{eqnarray}
  I_{1\bf k} &=& \frac{S}{16} \frac{J}{(2 \pi)^6}
	\int d {\bf k}_1 d {\bf k}_2 \frac{1-\cos(k_{1z}+k_{2z})}{\varepsilon_{\bf k} - \varepsilon_{{\bf k}_1} - \varepsilon_{{\bf k}_2}}, \\
  I_{2\bf k} &=& \frac{S J}{12}\sum_{\nu_{1},\nu_{2},\nu_3}  \frac{1}{(2\pi)^6}
		\int d {\bf k}_1 d {\bf k}_2
		\frac{ 1-\cos{(k_{1\nu_3}+k_{2\nu_3})}  }{(\varepsilon_{\bf k}-\varepsilon_{{\bf k}_1} - \varepsilon_{{\bf k}_2})(\tilde{\mathbf{k}}_1+\tilde{\mathbf{k}}_2)^4}\nonumber \\
		&& \times (1+\cos{(k_{1\nu_{1}}+k_{2\nu_{1}})}-\cos{k_{1\nu_{1}}}-\cos{k_{2\nu_{1}}})(1+\cos{(k_{1\nu_{2}}+k_{2\nu_{2}})}-\cos{k_{1\nu_{2}}}-\cos{k_{2\nu_{2}}}),\\
  I_{3\bf k}	&=& \frac{S J}{6}\sum_{\nu_1,\nu_2} \frac{1}{ (2 \pi)^6}
		\int d{\bf k}_1 d{\bf k}_2
		\frac{ 1+\cos(k_{1\nu_1}+k_{2\nu_1})-\cos k_{1\nu_1}- \cos k_{2 \nu_1} }{(\varepsilon_{\bf k} - \varepsilon_{{\bf k}_1} - \varepsilon_{{\bf k}_2})\left(\tilde{\mathbf{k}}_1+\tilde{\mathbf{k}}_2\right)^2}
		\sin^2{\frac{k_{1\nu_2}+k_{2\nu_2}}{2}},
\end{eqnarray}
where $\nu_{1,2,3}=x,y,z$. One can see from Eqs.~\eqref{bareb20} and \eqref{gamb20} that the damping is small compared to the bare spectrum, $\gamma_{\bf k} \ll \varepsilon^{(0)}_{\bf k}$. The correction to the magnon energy \eqref{deb20} is much larger than the bare spectrum \eqref{bareb20} for small enough momenta signifying a new physics at such $\bf k$. However, this effect is screened in real B20 materials at $c\ll1$ by the small gap in the bare spectrum mentioned above.

\section{Summary}
\label{sum}

To summarize, we develop a theory describing spiral magnets with bond disorder at small concentration $c$ of defects. It is assumed that both DMI and exchange coupling are changed on imperfect bonds. We obtain qualitatively the same physical picture in two models which are considered in detail: layered and B20 cubic helimagnets. Using the Holstein-Primakoff spin representation, we find the distortion of the spiral magnetic ordering around a single imperfect bond. It is shown that values of additional turns of spins caused by the impurity are given by Poisson's equation for electric dipole. Thus, the magnetic ordering distortion from a single imperfect bond is long-range: values of the additional turns of spins decay with the distance $r$ to the defect as $1/r^2$. Poisson's equations for the dipole in the corresponding models on lattices with space dimensions $d\ge2$ give the power-decaying law $1/r^{d-1}$.

At finite concentration of randomly distributed defect bonds, we calculate the observable quantities by averaging over disorder configurations. We find that the direction of the spiral vector does not change and its modulus acquires a correction $\delta q$ given by Eqs.~\eqref{dq} and \eqref{dqb20} in the two models considered. For defects of the type shown in Fig.~\ref{b20}(b) in cubic magnets, $\delta q$ is given by Eq.~\eqref{dqb20_2}. It is seen from these equations that the spiral vector correction can be zero, positive or negative depending on the particular parameters of defects. For negative $\delta q$, the sign of chirality can change even at $c\ll1$ if defects are strong enough.

In the elastic neutron scattering cross-section, defects manifest themselves in two ways. First, magnetic Bragg peaks (satellites) are shifted from reciprocal lattice vectors by $\pm({\bf q + \delta q})$ (i.e., by values defined by the new spiral vector). Second, diffuse scattering arises which has power-law singularities at Bragg peaks positions. Then, each Bragg peak acquires the power-law decaying tails (see Eqs.~\eqref{simpleCS}, \eqref{B20CS}, and Fig.~\ref{neutron}). This feature is attributed to the long-range character of the perturbation made by defect bonds.

Corrections to the magnon energy and to the damping caused by scattering on defects are given by Eqs.~\eqref{speccorrgen}--\eqref{gamcorrgen} and \eqref{deb20}--\eqref{gamb20} in layered and B20 magnets, respectively. The magnon damping is found to be much smaller than the bare spectra in both models. Although magnons are well defined at $k\gg q$ in both models, the ratio $\gamma_{\bf k}/\varepsilon_{\bf k}\sim c/k$ is quite unusually large. Remember, this ratio is normally proportional to a positive power of $k$ and it does not exceed $c$ in magnetically ordered gapless magnets (see, e.g., Refs.~\cite{wan,2dvac,syromyat1,syromyat2} and references therein). However we have obtained recently that $\gamma_{\bf k}/\varepsilon_{\bf k} \sim c/k^2$ under certain conditions in gapped phases of 3D spin systems with bond disorder. \cite{oleg}

Corrections to the magnon energy exceeds the bare spectra at small enough momenta. This signifies that the analysis cannot be restricted by the first order in defects concentration at such $k$. It can also signify a localization of long-wavelength magnons (see, e.g., Ref.\cite{oleg} and references therein). Consideration of this point is out of the scope of the present paper. Besides, these small-energy peculiarities can be screened by a small gap in the bare spectra originating from a small low-symmetry spin interaction.

Although all calculations for layered helimagnets are performed for the model with FM exchange interactions, the results obtained (except for the spectrum renormalization) are applicable after simple modifications discussed in Sec.~\ref{otherlay} to many other layered helimagnets with bond disorder.

Our consideration can be relevant to Mn$_{1-x}$Fe$_x$Ge considered recently experimentally in Ref.~\cite{grig2013}. But we are unable now to verify our theory due to very small amount of experimental data at $x\approx1$. For instance, there are only three experimental points on the plot for dependence of the spiral vector modulus on $x$ at $x>0.75$. Then, further experimental activity is needed in this field.

\begin{acknowledgments}

This work is supported by Russian Scientific Fund Grant No.\ 14-22-00281. One of us (O.I.U.) acknowledges the Dynasty foundation for partial financial support.

\end{acknowledgments}

\appendix

\section{Calculation of the magnon spectrum renormalization in layered spiral magnets}
\label{append}

First, we take into account the imperfection of the DMI only. In addition to terms in $\mathcal{V}_{dm}$ presented in Eq.~\eqref{DMpert}, one needs also the following terms for the magnon spectrum calculation:
\begin{eqnarray}
  \label{DMpert2}
\mathcal{V}^{(2)}_{dm} &=& S u_{dm} q
\sum_{hm}\left( a^+_{hm}a_{hm} + a^+_{hm+1}a_{hm+1} - \frac{1}{2} \left( a^+_{hm} a^+_{hm+1} + a_{hm} a_{hm+1} + a^+_{hm} a_{hm+1} + a^+_{hm+1} a_{hm}\right) \right),\\
  \label{DMpert3}
\mathcal{V}^{(3)}_{dm} &=& \sqrt{\frac{S}{2}} u_{dm}
\sum_{hm}\left( a^+_{hm} a_{hm} a_{hm+1} + a^+_{hm} a^+_{hm+1} a_{hm} + \frac{a^+_{hm+1} a^2_{hm+1}}{4} + \frac{a^{+2}_{hm+1} a_{hm+1}}{4} \right.\nonumber\\
&&{}-\left. a^+_{hm+1} a_{hm} a_{hm+1} - a^+_{hm} a^+_{hm+1} a_{hm+1} - \frac{a^+_{hm} a^2_{hm}}{4} - \frac{a^{+2}_{hm} a_{hm}}{4}\right),
\end{eqnarray}
where sums run over sites involved in defect bonds. Besides, one has to take into account terms in the Hamiltonian containing products of four Bose operators
\begin{eqnarray}
	\label{h4}
\mathcal{H}_4 &=& -J_0 \sum_{in} \left[ a^+_{in}a^+_{in+1} a_{in} a_{in+1} - \frac{1}{4} \bigl( a^{+2}_{in+1} a_{in} a_{in+1} + a^{+2}_{in} a_{in} a_{in+1}+a^+_{in}a^+_{in+1} a^2_{in+1} + a^+_{in}a^+_{in+1} a^2_{in}\bigr)\right]  \nonumber \\
  &&{}- J_1 \sum_{\langle ij\rangle n} \left[ a^+_{in}a^+_{jn} a_{in} a_{jn} - \frac{1}{2} \bigl( a^{+2}_{jn} a_{in} a_{jn} + a^+_{in}a^+_{jn} a^2_{jn} \bigr)\right],
\end{eqnarray}
where we omit terms of the second order in $D/J_0\ll1$.

Eq.~\eqref{h4} gives the following terms after shift \eqref{shift} which contain products of one operator of creation and one operator of annihilation:
\begin{equation}
\label{h42}
  \mathcal{H}^{(2)}_4 = -2\sum_{in} \sum_j J_j
\left[ b^+_{in} b_{in} (\rho^2_{jn}-\rho_{in}\rho_{jn})
-
\frac12 b^+_{in} b_{jn} (\rho_{in}-\rho_{jn})^2
\right],
\end{equation}
where $j$ enumerates nearest neighbors of the $i$-th site in the $n$-th plane, $J_j=J_0$ and $J_j=J_1$ for neighbors from different planes and from the same plane, respectively. It can be shown that terms containing products of two operators of creation or two operators of annihilation give a negligible correction to the spectrum. Introducing the Fourier transform
\begin{equation}
  b_{in}=\frac{1}{\sqrt{N}}\sum_{\mathbf{k}} b_{\bf k} e^{-i \mathbf{k}\cdot \mathbf{R}_{in}}
\end{equation}
we have for Eq.~\eqref{h42}
\begin{eqnarray}
    \label{Pert1}
    \mathcal{H}^{(2)}_4 &=&
		-\frac 2N
		\sum_{{\bf k}_1,{\bf k}_2} b^+_{{\bf k}_1} b_{{\bf k}_2}
		\sum_{in}\sum_j J_j e^{i \mathbf{R}_{in} \cdot (\mathbf{k}_1-\mathbf{k}_2)}
		\left[ \rho^2_{jn}-\rho_{in}\rho_{jn}
		-\frac12 \cos k_{2j} ( \rho_{in}-\rho_{jn} )^2
		\right],
\end{eqnarray}
where $k_{2j}=k_{2z}$ and $k_{2j}=k_{2x}$ or $k_{2y}$ for neighbors from different planes and from the same plane, respectively.

In much the same way, one obtains for terms containing products of three Bose-operators and stemming from $\mathcal{H}_4$ (Eq.~\eqref{h4})
\begin{equation}
  \label{Pert2}
  \mathcal{H}^{(3)}_4 =
	-\frac{1}{N^{3/2}}
	\sum_{{\bf k}_1, {\bf k}_2, {\bf k}_3}  b^+_{{\bf k}_1} b^+_{{\bf k}_2} b_{{\bf k}_3}
	\sum_{in}\sum_j J_j e^{i \mathbf{R}_{in}\cdot(\mathbf{k}_1+\mathbf{k}_2-\mathbf{k}_3)}
	\Bigl(e^{i k_{1j}}\left( 2\rho_{jn} - \rho_{in} \right)
	- e^{-i k_{3j}}\left(1+e^{i(k_{1j}+k_{2j})}\right)\frac{\rho_{in}}{2}\Bigr) + {\rm h.c.}
\end{equation}

Taking into account only terms in Eq.~\eqref{DMpert2} containing products of one operator of creation and one operator of annihilation which give the main contribution to the spectrum renormalization, one obtains
\begin{eqnarray}
  \label{Pert3}
  \mathcal{V}^{(2)}_{dm} = \frac{S q u_{dm}}{N} \sum_{{\bf k}_1, {\bf k}_2} b^+_{{\bf k}_1} b_{{\bf k}_2} \sum_{hm} e^{i \mathbf{R}_{hm}\cdot(\mathbf{k}_1-\mathbf{k}_2)}\left[ 1+e^{i(k_{1z}-k_{2z})} - \frac{e^{ik_{1z}}+e^{-i k_{2z}}}{2}\right].
\end{eqnarray}
We have from Eq.~\eqref{DMpert3} after the Fourier transformation
\begin{equation}
  \label{Pert4}
  \mathcal{V}^{(3)}_{dm} = \frac{ u_{dm}\sqrt{S/2}}{N^{3/2}} \sum_{{\bf k}_1, {\bf k}_2, {\bf k}_3} b^+_{{\bf k}_1} b^+_{{\bf k}_2} b_{{\bf k}_3}
	\sum_{hm} e^{i \mathbf{R}_{hm}\cdot(\mathbf{k}_1 + \mathbf{k}_2 -\mathbf{k}_3)} \left( \frac{(e^{i k_{1z}}+e^{i k_{2z}})(1-e^{-i k_{3z}})}{2} + \frac{e^{i(k_{1z}+k_{2z}-k_{3z})}-1}{4} \right) + {\rm h.c.}
\end{equation}

Let us start the spectrum calculation with Eq.~\eqref{Pert3}. As $q|u_{dm}|\ll J_{0,1}$, the main corrections to the magnon energy $\delta \varepsilon_{\bf k}$ and to the damping $\gamma_{\bf k}$ originate from diagrams shown in Fig.~\ref{diagrams}(a) and \ref{diagrams}(b), respectively, which give
\begin{eqnarray}
\label{de1}
  \delta \varepsilon^{(1)}_{\bf k} &=& S c q u_{dm} (2-\cos k_z),\\
	\gamma_{\bf k}^{(1)} &=& \Im \left( \frac{(S q u_{dm})^2}{N^2}
	\sum_{{\bf k}_1} \frac{1}{\varepsilon_{\bf k}-\varepsilon_{{\bf k}_1}-i0}
	\left| 1+ e^{i (k_z-k_{1z})} - \frac{\left( e^{ik_{z}}+e^{-i k_{1z}}\right)}{2} \right|^2
	\overline{\sum_{hm,h'm'} e^{i (\mathbf{R}_{hm}-\mathbf{R}_{h'm'})\cdot(\mathbf{k}_1-\mathbf{k})}}
	\right) \nonumber \\
	\label{gamma1}
	&\approx& \Im \left( c \frac{(S q u_{dm})^2}{(2\pi)^3} \int \frac{d^3 {\bf k}_1}{\varepsilon_{\bf k}-\varepsilon_{{\bf k}_1}-i0} \right)
	\approx c \frac{k^3}{\varepsilon_{\bf k}}\frac{(Su_{dm}D)^2}{J_0 J_1} \frac{t}{2\pi},
\end{eqnarray}
where $\Im$ denotes imaginary part, $t=1$ and $1/2$ for, respectively, $\tilde{k} \ll D/J_0$ and $\tilde{k} \gg D/J_0$ (see Eqs.~\eqref{spec0}--\eqref{reg2}), hereafter the line over an expression denotes averaging over disorder configurations, and we take into account that only terms with $h=h'$ and $m=m'$ survive after the averaging over disorder configurations in the double sum over $hm$ and $h'm'$.

The main contribution to the spectrum renormalization from Eq.~\eqref{Pert4} originates from the diagram presented in Fig.~\ref{diagrams}(c). After integration over internal frequency and averaging over disorder configurations, we have for it
\begin{equation}
  cu^2_{dm}\frac{S}{8N^2}
	\sum_{{\bf k}_1,{\bf k}_2}
	\frac{1}{\varepsilon_{\bf k} -\varepsilon_{{\bf k}_1} - \varepsilon_{{\bf k}_2} -i0 }
	\left| (e^{i k_{1z}}+e^{i k_{2z}})(1-e^{-i k_{z}}) + \frac{e^{i(k_{1z}+k_{2z}-k_{z})}-1}{2} \right|^2.
\end{equation}
The imaginary part of this equation is of the order of $c u^2_{dm} k^{10}/\varepsilon_{\bf k}$. Thus, it is larger than Eq.~\eqref{gamma1} only for quite large momenta, $k\gg(D/\sqrt{J_0J_1})^{2/7}$. The correction to the magnon energy has the form
\begin{eqnarray}
\label{d32}
  \delta \varepsilon^{(2)}_{\bf k}
		&\approx& c \frac{u^2_{dm}}{J_0} I_{1\bf k}, \\
\label{i1}
		I_{1\bf k} &=& \frac{S}{16} \frac{J_0}{(2 \pi)^6}
	\int d {\bf k}_1 d {\bf k}_2 \frac{1-\cos(k_{1z}+k_{2z})}{\varepsilon_{\bf k} - \varepsilon_{{\bf k}_1} - \varepsilon_{{\bf k}_2}}
\end{eqnarray}
that should be taken into account together with Eq.~\eqref{de1}.

The main correction to the magnon energy from Eq.~\eqref{Pert1} is given by the diagram shown in Fig.~\ref{diagrams}(a):
\begin{equation}
	\overline{\sum_{in}\sum_j J_j (\cos k_j-1)
	 (\rho_{in} - \rho_{jn})^2 }.
\end{equation}
It is negligible compared to Eq.~\eqref{d32} being of the order of $cu_{dm}^2k^2$. Contribution to the damping from Eq.~\eqref{Pert1} stems from the diagram depicted in Fig.~\ref{diagrams}(b) and it has the form
\begin{eqnarray}
\label{gamma2}
\gamma_{\bf k}^{(2)} &=&   \Im \left(\frac{1}{N^2} \sum_{{\bf k}_1}
	\frac{1}{\varepsilon_{\bf k} -\varepsilon_{{\bf k}_1}-i0}
	\overline{\sum_{in}\sum_{i'n'} \sum_{j,j'}J_{j}J_{j'}
	e^{i (\mathbf{R}_{in}-\mathbf{R}_{i'n'})\cdot(\mathbf{k}_1-\mathbf{k})}
 \left(\rho_{in}^2 - \rho_{jn}^2\right) \left(\rho_{i'n'}^2 - \rho_{j'n'}^2\right)}\right).
\end{eqnarray}
It can be discarded being of the order of $cu_{dm}^4k^5/\varepsilon_{\bf k}$.

The loop diagram shown in Fig.~\ref{diagrams}(c) with three-particle vertex \eqref{Pert2} gives
\begin{eqnarray}
\label{expert2}
	&&\sum_{j_{1},j_{2}} \frac{J_{j_{1}}J_{j_{2}}}{4N^3}
	\sum_{{\bf k}_1,{\bf k}_2} \frac{(1+\cos{(k_{1j_{1}}+k_{2j_{1}})}-\cos{k_{1j_{1}}}-\cos{k_{2j_{1}}})(1+\cos{(k_{1j_{2}}+k_{2j_{2}})}-\cos{k_{1j_{2}}}-\cos{k_{2j_{2}}})}{\varepsilon_{\bf k}-\varepsilon_{{\bf k}_1} - \varepsilon_{{\bf k}_2}-i0} \nonumber \\
	&&\times \overline{ \sum_{in,jm} \rho_{in} \rho_{jm} e^{i (\mathbf{R}_{in}-\mathbf{R}_{jm})\cdot(\mathbf{k}_1+\mathbf{k}_2)} }.
\end{eqnarray}
Using Eq.~\eqref{DipoleFourier}, one finds the following valuable contribution to the magnon energy from Eq.~\eqref{expert2}:
\begin{eqnarray}
    \delta \varepsilon^{(3)}_{\bf k} &=&
		c \frac{u_{dm}^2}{J_0}I_{2\bf k},\\
\label{i2}
		I_{2\bf k} &=& \sum_{j_{1},j_{2}} \frac{S J_{j_1}J_{j_2}}{4J_0} \frac{1}{(2\pi)^6}
		\int d {\bf k}_1 d {\bf k}_2
		\frac{ 1-\cos{(k_{1z}+k_{2z})}  }{(\varepsilon_{\bf k}-\varepsilon_{{\bf k}_1} - \varepsilon_{{\bf k}_2})(\tilde{\mathbf{k}}_1+\tilde{\mathbf{k}}_2)^4}\nonumber \\
		&& \times (1+\cos{(k_{1j_{1}}+k_{2j_{1}})}-\cos{k_{1j_{1}}}-\cos{k_{2j_{1}}})(1+\cos{(k_{1j_{2}}+k_{2j_{2}})}-\cos{k_{1j_{2}}}-\cos{k_{2j_{2}}}),
\end{eqnarray}
which is of the order of $c u_{dm}^2$ as Eq.~\eqref{d32}. The imaginary part of Eq.~\eqref{expert2} is of the order of $c u^2_{dm} k^{8}/\varepsilon_{\bf k}$. Thus, it is larger than Eq.~\eqref{gamma1} only for quite large momenta, $k\gg(D/\sqrt{J_0J_1})^{2/5}$.

There are also corrections from the diagram shown in Fig.~\ref{diagrams}(b) which is built using both ${\cal V}^{(2)}_{dm}$ and $\mathcal{H}^{(2)}_4$. The corresponding expression has the form
\begin{eqnarray}
    &&-\sum_\nu \frac{J_\nu q u_{dm} S}{N^2}
		\sum_{{\bf k}_1} \frac{1}{\varepsilon_{\bf k} -\varepsilon_{{\bf k}_1}-i0}
		\Bigl( \sum_{\{ in \} } e^{i \mathbf{R}_{in}\cdot(\mathbf{k}_1-\mathbf{k})}\left[ 1+e^{i(k_{1z}-k_{z})} - \frac{\left( e^{ik_{1z}}+e^{-i k_{z}}\right)}{2}\right] \nonumber \\
    &&\times\sum_{jm}  e^{i \mathbf{R}_{jm} \cdot (\mathbf{k}-\mathbf{k}_1)} \Bigl[ \rho^2_{jm+e_\nu}-\rho_{jm}\rho_{jm+e_\nu}- \rho_{jm-e_\nu}\rho_{jm}  +  e^{-ik_{1\nu}}\left(\rho_{jm}\rho_{jm+e_\nu}-\frac{\rho^2_{jm+e_\nu}+\rho^2_{jm}}{2}\right) \nonumber  \\
		&&+ e^{ik_{1\nu}}\left(\rho_{jm}\rho_{jm-e_\nu}-\frac{\rho^2_{jm-e_\nu}+\rho^2_{jm}}{2}\right)\Bigr] + {\rm h.c.} \Bigr).
\end{eqnarray}
Corrections to the magnon energy and to the damping from this expression are negligible being of the order of $c u^3_{dm} D$ and $c u_{dm}^3 D k^5/\varepsilon_{\bf k}$, respectively.

The second correction of this type comes from the loop diagram presented in Fig.~\ref{diagrams}(c)  which contains both vertexes ${\cal V}^{(3)}_{dm}$ and $\mathcal{H}^{(3)}_4$ and has the form
\begin{eqnarray}
    &&\sum_\nu \frac{ \sqrt{\frac{S}{2}}u_{dm} J_\nu}{N^3} \sum_{k_1,k_2}\frac{1}{\omega -\varepsilon_{k_1} - \varepsilon_{k_2} +i0 } \Bigl[ \sum_{\{in\}} e^{i \mathbf{R}_{in}\cdot(\mathbf{k} -\mathbf{k}_1 -\mathbf{k}_2)} \left( \frac{(e^{-i k_{1z}}+e^{-i k_{2z}})(1-e^{i k_{z}})}{2} + \frac{e^{i(k_{z}-k_{1z}-k_{2z})}-1}{4} \right)\nonumber \\
    &&\times \sum_{jm} e^{i \mathbf{R}_{in}\cdot(\mathbf{k}_1 +\mathbf{k}_2 -\mathbf{k})}\Bigl( \rho_{jm} \left[ \left(e^{-i k_\nu}- \frac{1}{2}\right) \left( \frac{e^{i k_{1\nu}}+e^{i k_{2\nu}}}{2}\right) -\frac{e^{- i k_\nu}}{4}(1+e^{i(k_{1\nu}+k_{2\nu})})\right] \\
		&& + \rho_{jm+e_\nu} \left[ \left(1 - \frac{ e^{-i k_\nu}}{2}\right) \left( \frac{e^{i k_{1\nu}}+e^{i k_{2\nu}}}{2}\right) -\frac{1}{4}(1+e^{i(k_{1\nu}+k_{2\nu})})\right] \Bigr) + {\rm h.c.} \Bigr]. \nonumber
\end{eqnarray}
The imaginary part of this expression at small $k$ is of the order of $c u^2_{dm} k^8/\varepsilon_{\bf k}$. Then, it is negligibly small. The real part is given by the following equation:
\begin{eqnarray}
    \delta \varepsilon^{(4)}_{\bf k}
    &=& c\frac{ u_{dm}^2}{J_1} I_{3\bf k},\\
\label{i3}
	I_{3\bf k}	&=& \sum_\nu \frac{SJ_\nu}{2 (2 \pi)^6}
		\int d{\bf k}_1 d{\bf k}_2
		\frac{ 1+\cos(k_{1\nu}+k_{2\nu})-\cos k_{1\nu}- \cos k_{2 \nu} }{(\varepsilon_{\bf k} - \varepsilon_{{\bf k}_1} - \varepsilon_{{\bf k}_2})\left(\tilde{\mathbf{k}}_1+\tilde{\mathbf{k}}_2\right)^2}
		\sin^2{\frac{k_{1z}+k_{2z}}{2}}
\end{eqnarray}
which is of the order of $c u_{dm}^2$ and should be taken into account.

%To summarize, correction to the magnon energy and the damping from the defect in the DMI only is negligibly small compared to the bare spectrum being of the order of $c D u_{dm} + c u^2_{dm}$ and $c D^2 u^2_{dm} k^3/\varepsilon_{\bf k}$, respectively.

Let us take into account the defect in the exchange interaction \eqref{FMpert} which has the following form after the Fourier transformation:
\begin{equation}
  \label{FMpert2}
  \mathcal{V}^{(2)}_{ex} = S u_{ex} \sum_{\{in\}} \sum_{{\bf k}_1,{\bf k}_2} \frac{1}{N} b^+_{{\bf k}_1} b_{{\bf k}_2} e^{{i \mathbf{R}_{in}\cdot(\mathbf{k}_1-\mathbf{k}_2})}
	\left(1-e^{i k_{1z}}\right)
	\left(1-e^{-i k_{2z}}\right),
\end{equation}
where $\{in\}$ denotes imperfect bonds. In general, one cannot assume that $|u_{ex}|\ll J_0,J_1$ as it was for $u_{dm}$. Then, one has to sum an infinite set of diagrams of the type shown in Fig.~\ref{diagrams}(d) to find spectrum corrections in the first order in $c$ from Eq.~\eqref{FMpert2}. As a result, the Green's function denominator has the form
\begin{eqnarray}
  G(\omega, {\bf k})^{-1} &=& \omega-\varepsilon^{(0)}_{\bf k}-T(\omega, {\bf k}),\\
  T(\omega, {\bf k}) &=& c 2 S u_{ex} (1-\cos k_z)
	\left(
	1- 2 S u_{ex} \int \frac{d {\bf q}}{(2\pi)^3} \frac{(1-\cos q_z)}{\omega -\varepsilon^{(0)}_{\bf q} -i 0 }
	\right)^{-1}.
\end{eqnarray}
One has from this expressions
\begin{eqnarray}
  \gamma_{\bf k} &=& c\frac{S u_{ex}^2 k^6}{6 \sqrt{2} \pi^2 D (1 + I_4 u_{ex})^2}, \\
  \delta \varepsilon_{\bf k} &=& c \frac{S u_{ex} k^2}{1 + I_4 u_{ex}},\\
	I_4 &=& 2S\int \frac{d{\bf q}}{(2\pi)^3} \frac{1-\cos q_z}{\varepsilon^{(0)}_{\bf q} }.
\end{eqnarray}
These results are negligible compared with those stemming from the defect in DMI which are considered above. There are also corrections from diagrams of the type Fig.~\ref{diagrams}(d) made both from Eqs.~\eqref{Pert3} and \eqref{FMpert2}. Their analysis shows that they are also small.

%Thus, in the case of nonzero $u_{ex}$, we need to consider only small (proportional to some power of $q$) corrections to $V^{(2)}$ and $V^{(3)}$, remembering that the dipoles charge is given by Eq.~\eqref{qres} which contains $u_{ex}$ as well. This leads to obvious changes in corresponding equations.

\bibliography{bibliography}

\end{document}